\documentclass[12pt,a4paper]{article}
\usepackage{amsmath}
\usepackage{amsxtra}
    \usepackage{amstext}
    \usepackage{amssymb}
    \usepackage{latexsym}
    \usepackage{graphicx}
\usepackage{color}
\usepackage{graphics}
\usepackage{cite}

\newcommand{\cm}{{\cal M}}

\newcommand{\ce}{{\cal E}}
\newcommand{\cp}{{\cal P}}
\newcommand{\cd}{{\cal D}}

\newcommand{\p}{\vspace{6pt}\noindent}
\newcommand{\jump}{\vspace{2pt}}

\advance\textheight by \topskip
\makeatletter

\@addtoreset{equation}{section}
\def\section{\@startsection {section}{1}{\z@}{-8.5ex plus -1ex minus
 -.2ex}{3.3ex plus .2ex}{\large\bf}}
\def\subsection{\@startsection{subsection}{2}{\z@}{-3.25ex plus
 -1ex minus -.2ex}{1.5ex plus .2ex}{\bf}}
\def\subsubsection{\@startsection{subsubsection}{3}{\z@}{-3.25ex plus%
 -1ex minus -.2ex}{1.5ex plus .2ex}{\sl}}

\begin{document}

\begin{titlepage}
\vspace*{-2cm}
\begin{flushright}
\end{flushright}

\vspace{0.3cm}

\begin{center}
{\Large {\bf Integrable Defects at Junctions within a Network}} \\
\vspace{1cm} {\large  E.\ Corrigan\footnote{\noindent E-mail: {\tt edward.corrigan@york.ac.uk}}}\\
\vspace{0.5cm}
{\em Department of Mathematics \\ University of York, York YO10 5DD, U.K.} \\
\vspace{0.3cm} {\large and}\\ \vspace{0.5cm}
{\large C.\ Zambon\footnote{\noindent E-mail: {\tt cristina.zambon@durham.ac.uk}}} \\
\vspace{0.3cm}
{\em Department of Physics \\ Durham University, Durham DH1 3LE, U.K.} \\

\vspace{2cm} {\bf{ABSTRACT}}\\ \end{center}   \vspace{.5cm}

\begin{center}
\begin{minipage}[]{14cm}
  The purpose of this article is to explore the properties of integrable, purely transmitting, defects placed at the junctions of several one-dimensional domains within a network. The defect sewing conditions turn out to be quite restrictive - for example, requiring the number of domains meeting at a junction to be even - and there is a clear distinction between the behaviour of conformal and massive integrable models. The ideas are mainly developed within classical field theory and  illustrated using a variety of  field theory models defined on the branches of the network, including both linear and nonlinear examples.
\end{minipage}
\end{center}

\vfill
\end{titlepage}

\section{Introduction}

Defects within (relativistic) integrable field theory models in two dimensions have been studied for some time from both classical and quantum viewpoints (see, for example \cite{dms1994,kl1999,Mintchev02,bcz2003,bcz2004, bcz2005,cz2005,gyz2006,cz2007,hk2007,gyz2007,c2008, cz2009,cz2010, n2009,ad2012,agsz2014,d2016,pd2019}). In addition, some non-relativistic systems, for example the nonlinear Schr\"odinger, KdV and mKdV equations have been shown to support defects classically \cite{cz2005,gyz2006,c2008 }. In essence, a defect always involves a discontinuity of some kind, and in an integrable model experience has shown that this discontinuity is a `jump' in the field value at a specific point (similar to the discontinuity in velocity across a shock in a fluid flow), with `sewing' conditions across the defect relating the fields on either side in such a manner that suitably adjusted energy and momentum conservation laws are maintained. At least in the relativistic case, presenting the argument the other way round, the presence of defects that preserve a suitably modified energy and momentum seems to require the fields on either side of the defect to be integrable. Characteristically, such defects break space translation invariance (since they have a specific location) but are purely transmitting. In this sense, they are distinct from `delta function' type discontinuities that typically involve reflection as well as transmission.

\p So far, there are basically two types of defect that appear to be integrable, called type I (where the defect has no degrees of freedom of its own \cite{bcz2003,bcz2004}), and type II (where the defect carries its own degrees of freedom \cite{cz2009,cr2013}). Both types are needed to discuss defects within the $d^{(1)}_r$ series of affine Toda field theories \cite{br2017}. There may be other possibilities, yet to be found, that for example encompass affine Toda field theories based on the $e_r^{(1)},\ r=6,7,8$ root systems, or extensions to higher dimensional integrable systems such as the Kadomtsev-Petviashvili equation \cite{konopelchenko92}.

\p In a  series of articles \cite{sobirov2010,sobirov2015} it was suggested that nonlinear wave equations can be defined on graphs or networks by developing suitable junction conditions (but see  a much earlier paper  \cite{nakajima1976}, where similar ideas are developed in a different context, and also \cite{caputo2014}). In particular, the sine-Gordon model can be adapted in a manner suitable for a network by allowing the basic wave speeds to be different within its separated segments \cite{sobirov2015}. Characteristic of these particular junction conditions is continuity of the fields as they match at a junction.

\p In this article, one aim is to demonstrate the existence of defects  that join just two domains with different wave speeds. For this to be possible there must be a discontinuity. Another, is to explore the possibilities for constructing defects at  junctions linking more than two domains while continuing to preserve conserved energy and momentum. One reason for investigating these features is the similarity the `jump' defects bear to junctions in nerve fibres (synapses), see \cite{katz1966}; these too represent a discontinuity and are (ideally) purely transmitting. The articles \cite{heimberg2005,lautrup2011,appali2012} provide detailed discussions of solitons on nerves, though the soliton-supporting wave equations considered there are not yet known to support defects of the type discussed here - investigating that is future task. Another reason might be to devise a mechanism for moving solitons around in two or three dimensions by using junctions joining at least four or six branches, respectively, which do not split solitons, but act as switches between the various branches. Because the junctions are defects they can, with sine-Gordon models defined on each branch, store topological charge as a soliton converts (under certain circumstances \cite{bcz2005}) to an antisoliton, or is removed from the network by being stopped at a junction. In the latter sense, the present paper might be considered a supplement to the speculative paper \cite{cz2004}, which introduced the idea  of using a combination of solitons and a defect to construct a model universal Toffoli gate \cite{Toffoli80}.

\section{Integrable defects}

\p In this section integrable defects are reviewed briefly and adapted to join two domains with different wave speeds. This is a relatively straightforward development that also serves to set the scene for future sections. Note, in the next subsection the subscripts $I$ and $II$ will be used to emphasise the focus on \lq type I' or  \lq type II' defects. However, in the remaining part of the article it will be made clear which kind of defect or junction is the subject of the investigation and, as a consequence, these subscripts will be abandoned.

\subsection{The formalism}

\p It is useful to begin with a Lagrangian description of a sine-Gordon defect  located at $x_0$ joining two one-dimensional domains (with fields labelled $u(x,t),\ {\rm for}\ x<x_0$ and $v(x,t),\ {\rm for}\ x>x_0$). Thus:
\begin{equation}\label{defectL}
{\cal L}(u,v)={\cal L}(u)\theta(x_0-x)+{\cal L}_D\delta(x-x_0)+{\cal L}(v)\theta(x-x_0),
\end{equation}
with
\begin{equation}\label{defectLI}
{\cal L}(u)=\frac{1}{2}(u_t^2-c_u^2 u_x^2)-\frac{m_u^2}{\beta_u^2}(1-\cos(\beta_u u)),\quad {\cal L}(v)=\frac{1}{2}( v_t^2-c_v^2 v_x^2) -\frac{m_v^2}{\beta_v^2}(1-\cos(\beta_v v)),
\end{equation}
and with the type I or type II defect Lagrangian ${\cal L}_D$ given by
\begin{equation}\label{defectLII}
{\cal L}_I=\sqrt{c_uc_v}\,uv_t-{\cal D}_I(u,v),\quad {\cal L}_{II}=(\sqrt{c_u}\,u-\sqrt{c_v}\,v)\lambda_t-{\cal D}_{II}(u,v,\lambda).
\end{equation}
In these expressions, subscripts $t,x$ denote derivatives with respect to $t$ and $x$,  and the defect energy functional ${\cal E}$ depends only on the fields not their time (or space) derivatives. Because the media are different for $x<x_0$ and $x>x_0$, in the sense that the wave speeds are $c_u,\, c_v$, respectively, it is also necessary to pay attention to the other constants in the Lagrangians that might be different on either side of the defect. Also, the fields $u,\, v$ are evaluated at $x_0$ in a limiting sense,
$$u(x_0,t)=\lim_{\epsilon\rightarrow 0}\, u(x_0-\epsilon),\  \  v(x_0,t)=\lim_{\epsilon\rightarrow 0}\, v(x_0+\epsilon),$$
where $\epsilon>0$.

\p Using these Lagrangians, the sewing conditions across a defect have the form
\begin{equation}\label{sewingI}
c_u^2 u_x=\sqrt{c_uc_v}\,v_t-\frac{\partial{\cal D}_I}{\partial u},\ \ c_v^2 v_x=\sqrt{c_uc_v}\,u_t+\frac{\partial{\cal D}_I}{\partial v},
\end{equation}
for the type I case, and
\begin{equation}\label{sewingII}
c_u^2 u_x=\sqrt{c_u}\,\lambda_t-\frac{\partial{\cal D}_{II}}{\partial u},\ \ c_v^2 v_x=\sqrt{c_v}\,\lambda_t+\frac{\partial{\cal D}_{II}}{\partial v},\ \ \sqrt{c_u}\,u_t-\sqrt{c_v}\,v_t=-\frac{\partial{\cal D}_{II}}{\partial\lambda},
\end{equation}
for type II. The defect contribution in either case will be derived below but the type I case   is provided here as an illustration:
\begin{equation}\label{DIsineGordon}
{\cal D}_I(u,v)=\kappa m\left(\sigma\cos\frac{1}{2}(\beta_u u+\beta_v v)+\frac{1}{\sigma}\cos\frac{1}{2}(\beta_u u-\beta_v v)\right),
\end{equation}
where $\sigma\equiv \exp(-\eta)$ is a dimensionless parameter and
\begin{equation}\label{kappaequation}m_u=m_v\equiv m,\quad \kappa=\frac{c_u}{\beta_u^2}=\frac{c_v}{\beta_v^2}.\end{equation}
The latter requirement is quite strong and has other consequences.

\p For future reference, a sine-Gordon field $w$ in a domain with wave speed $c$  satisfies the following equation of motion
\begin{equation}\label{sG}
w_{tt}-c^2w_{xx}=-\frac{m^2}{\beta}\sin(\beta w),
\end{equation}
with a soliton solution given by
\begin{equation}\label{sGsoliton}
  e^{i\beta w/2}=\frac{1+iE}{1-iE}, \quad E=e^{ax+bt+d},\quad a=\frac{m\cosh\theta}{c},\quad b=-m\sinh\theta.
\end{equation}
It is already to be expected from the form of this solution that if a soliton  travelling in the direction of the positive $x$-axis (ie from $x<x_0$ to $x>x_0$) can be transmitted through a defect then on each side of the defect, for a successful matching via sewing conditions, the time dependence of the fields (in a limiting sense from either side of the defect) must be the same at the defect location, which implies that $m_u=m_v$ must be a requirement. Indeed, a soliton transmitted through a type I defect will be given by expressions of this kind valid on either side of the defect. Thus,
\begin{eqnarray}
 &&\phantom{mmm} e^{i\beta_u u/2}=\frac{1+iE_u}{1-iE_u},\  e^{i\beta_v v/2}=\frac{1+izE_v}{1-izE_v}, \nonumber\\ &&\nonumber\\
 &&a_u=\frac{m\cosh\theta}{c_u}, \ \  a_v=\frac{m\cosh\theta}{c_v},\ \  b_u=b_v =-m\sinh\theta,
\end{eqnarray}
where
\begin{equation}
z=\exp\left(\left(\frac{1}{c_u}-\frac{1}{c_v}\right)mx_0\cosh\theta\right)\coth\left(\frac{\eta-\theta}{2}\right).
\end{equation}
If $c_u=c_v$ the familiar result \cite{bcz2005} is recovered but it is interesting to observe that the \lq delay' when $c_u\ne c_v$ depends on the location of the defect at $x=x_0$. As previously \cite{bcz2005}, the soliton may emerge as a soliton if $\theta<\eta$, or flip to an anti-soliton if $\theta>\eta$, or be captured by the defect if its rapidity satisfies $\theta=\eta$. Note, the soliton speeds in the two regions are not the same: for $x<x_0$ the soliton is travelling at a speed $c_u\tanh\theta$ but for $x>x_0$ its speed is $c_v\tanh\theta$. If $x$ is replaced by $x-x_0$ in the expression \eqref{sGsoliton}, the extra factor in $z$ cancels out though the dependence on $x_0$ remains explicit in the solution. This was not the case previously when $c_u=c_v$.

 \subsection{Energy}
 \label{Energy}

 A characteristic of integrable defects is that they are defined by requiring energy-momentum to be preserved, which generally requires a contribution from the defect itself to both energy and momentum \cite{bcz2003, bcz2004}.
Thus, the total energy is given by
\begin{equation}\label{energy}
{\cal E}={\cal E}(u)+{\cal D}(u,v)+{\cal E}(v),
\end {equation}
where $\cd$ is a functional of the fields defined in a limiting sense, as described above, at the point $x=x_0$, with
\begin{eqnarray}
 \nonumber  && {\cal E}(u)=\int_{-\infty}^{x_0} \left(\frac{1}{2}(u_t^2+c_u^2 u_x^2)+\frac{m_u^2}{\beta_u^2}(1-\cos(\beta_u u))\right)dx, \\
 &&\\
\nonumber   && {\cal E}(v)
=\int^{\infty}_{x_0} \left(\frac{1}{2}(v_t^2+c_v^2 v_x^2)+\frac{m_v^2}{\beta_v^2}(1-\cos(\beta_v v))\right)dx
\end{eqnarray}
and the sewing conditions \eqref{sewingI}, \eqref{sewingII} guarantee it is conserved. Using the wave equation to either side of the defect it is straightforward to check that
$$\frac{d{\cal E}}{dt}=\left[c_u^2u_xu_t\right]^{x_0} +\frac{d{\cal D}}{dt}+\left[c_v^2v_xv_t\right]_{x_0}=0.$$
On the other hand, insisting that momentum is also conserved in either context places strong constraints on the defect contribution ${\cal D}$.

\subsection{Momentum}
\label{Momentum}

\p To ensure energy and momentum have the same dimensions,  consider the pair $({\cal E},\ cP)$. Then, the total contribution of the fields to the suitably scaled momentum will be taken to be
$${\cm}=\int_{-\infty}^{x_0} c_uu_tu_x dx +\int_{x_0}^\infty c_vv_tv_x dx.$$

\p In a similar manner, the time derivative of the contributions to the total field momentum is given by
\begin{eqnarray}\dot {\cm}=&&\int_{-\infty}^{x_0}\left(c_u u_t u_x\right)_t dx +\int_{x_0}^{\infty}\left(c_v v_t v_x\right)_t dx\nonumber\\&&\phantom{mmm}=\left[\frac{c_u}{2}(u_t^2+c_u^2u_x^2)-U(u)\right]^{x0}+\left[\frac{c_v}{2}(v_t^2+c_v^2 v_x^2)-V(v)\right]_{x_0},\end{eqnarray}
with the same assumption as before. Since space translation is broken explicitly by the defect the requirement of overall momentum conservation is expected to impose stringent conditions on the fields. The two cases introduced above will be dealt with separately.
\subsubsection{Type I}

\p Using the type I sewing conditions (in this section all fields are evaluated at $x=x_0$):
$$\dot {\cm}=-\sqrt{\frac{c_v}{c_u}}\,v_t\cd_u-\sqrt{\frac{c_u}{c_v}}\,u_t\cd_v +\frac{1}{2c_u}\cd_u^2-\frac{1}{2c_v}\ce_v^2 -c_uU(u)+c_vV(v)= -\frac{d\cp}{dt},$$
where $\cp$ is related to $\cd$ and strongly constrained by the following relationships:
\begin{equation}\label{typeIconditions}\sqrt{\frac{c_v}{c_u}}\,\cd_u=\cp_v,\ \ \sqrt{\frac{c_u}{c_v}}\,\cd_v=\cp_u,\ \ \frac{1}{2c_u}\,\cd_u^2-\frac{1}{2c_v}\,\cd_v^2 =c_u U(u)-c_v V(v).\end{equation}
The first pair of relations in \eqref{typeIconditions} then implies
$$\frac{1}{c_u}\,\cd_{uu}=\frac{1}{c_v}\,\cd_{vv}$$
while the second provides a nonlinear constraint on the solutions to this differential equation. As noted some years ago \cite{bcz2004}, there are very few solutions that can work, corresponding to massive free fields, massless free fields, Liouville fields and sine-Gordon fields. In the case of sine-Gordon, the appropriate solution for $\cd$ is given in \eqref{DIsineGordon}.

\subsubsection{Type II}

In this case, it is useful to define a pair of alternative variables at the defect point:
\begin{equation}\label{qpdefs}
q=\frac{1}{2}\left(\sqrt{c_u}\,u-\sqrt{c_v}\,v\right), \quad p=\frac{1}{2}\left(\sqrt{c_u}\,u+\sqrt{c_v}\,v\right),
\end{equation}
then the defect contribution to the energy-momentum depends on $q,p,\lambda$, where $\lambda$ is the additional degree of freedom carried by the defect. Requiring the total momentum
$$\cm=c_uP(u)+\cp(\lambda,p,q)+c_vP(v)$$ to be conserved requires
\begin{equation}\label{Poisson}\cd_p=\cp_\lambda,\ \ \ \cd_\lambda =\cp_p,\ \ \ \frac{1}{2}\left(\cd_\lambda\cp_q-\cd_q\cp_\lambda\right)=c_uU-c_vV.\end{equation}
As noted previously \cite{cz2009}, the third of these relations is a Poisson bracket with $\lambda,q$ as conjugate variables. It provides a powerful constraint  \cite{cz2018} because there is no dependence on $\lambda$ in the expression on the right hand side of \eqref{Poisson}.

\subsection{Remarks}
\label{Remarks}

It has been found that a energy-momentum preserving defect can be constructed between sine-Gordon field theories with scalar fields $u,v$ in two different media provided
\begin{equation}\label{parameterrelations}
m_u=m_v,\quad c_u\ne c_v,\ \ \ \beta_u\ne\beta_v, \quad \frac{c_u}{\beta_u^2}=\frac{c_v}{\beta_v^2}.
\end{equation}
Thus, after quantization (assuming $\hbar$ is universal), the dimensionless quantities defined by the field theory constants within the two media are given by
\begin{equation}\label{dimensionless}
\frac{c_u}{\hbar \beta_u^2}=\frac{c_v}{\hbar \beta_v^2}.
\end{equation}
Since this is the combination of constants on which the Zamolodchikov S-matrix depends \cite{Zam79}, it appears the two S-matrices should be identical, independently of the medium. This is something of a special case, of course, required by insisting the media are connected by an `integrable' defect.

\p Note also that after quantization the mass scales in the different media are not quite the same since $\hbar m/c^2$ has the dimension of mass and this is not the same in each medium, since $c_u\ne c_v$. In similar manner, the mass of a soliton is a classical feature and in the sine-Gordon model it is proportional to $m/c\beta^2$, which is also different in the two media. However, as a consequence of the second requirement of \eqref{parameterrelations} the energy-momentum carried by a soliton of a given rapidity will be the same on either side of the defect,
$$({\cal E}, {\cal M})=\frac{8\,c\,m}{\beta^2}\left(\cosh\theta,-\sinh\theta \right).$$

\section{Type I junction within a network}

\p In this section, the possibility of constructing a type I defect at a junction is investigated.

\subsection{The setting}
\label{Junctions on a graph}

A defect between two media can be thought of as a two-branch junction and in this section the aim is to generalise this idea in order to see if multi-branch junctions and defects can be combined. To begin with, the one-dimensional branches will be considered to meet at the common point $x_i=x_0$, where $x_i,\ i=1,\dots,N,$ are the spatial variables along the branches. Since the branches meet at a single point and are otherwise independent, the notation can be simplified, without causing confusion, by using the generic space variable $x$ along each branch. In what follows, since only a single junction is considered, $x_0=0$ is a convenient choice. Thus, the total energy and momentum are given by:
\begin{equation}\label{junctionEP}
\ce=\sum_{i=1}^N\epsilon_i\int_0^\infty\left(\frac{1}{2}(u^{(i)}_t) ^2+\frac{c_i^2}{2}(u^{(i)}_x)^2+U^{(i)}\right)dx,\quad \cm=\sum_{i=1}^N\epsilon_ic_i\int_0^\infty\left(u^{(i)}_tu^{(i)}_x\right)dx,
\end{equation}
where $\epsilon_i=\pm1,\ i=1,\dots,N,$ is introduced to take into account the sense of integration along a branch. For example, for two branches (which, as noted above, is a defect), $\epsilon_1=-\epsilon_2$, since the junction lies at the intersection of $[-\infty,0]$ and $[0,\infty]$). It is also convenient to define two diagonal matrices $c,\ \epsilon$ by
$$c={\rm diag}(c_i),\quad \epsilon={\rm diag}(\epsilon_i).$$
The search for junction sewing conditions is then a  generalisation of the arguments summarised in sections \ref{Energy} and \ref{Momentum}.

\p First, to conserve energy, it is enough to take junction conditions (at $x=0$):
\begin{equation}\label{junctionsewing}
c_i^2u^{(i)}_x=\epsilon_i\left(\sum_{j=1}^NA_{ij}u^{(j)}_t-\frac{\partial\cd\phantom{^{(i)}}}{\partial u^{(i)}}\right),\quad A^T=-A,\quad i=1,\dots,N,\end{equation}
where the `junction potential ${\cal D}$' is presumed to depend only on the fields, not on their derivatives. Also, because the fields have been assumed real, the matrix $A$ is also real. Then,
\begin{equation}\label{Edot}\dot\ce=\sum_{i=1}^N \epsilon_ic_i^2\left[u^{(i)}_tu^{(i)}_x\right]_{0}=\left[\sum_{i,j=1}^Nu^{(i)}_tA_{ij}u^{(j)}_t-\sum_{i=1}^Nu^{(i)}_t
\frac{\partial\cd\phantom{^{(i)}}}{\partial u^{(i)}}\right]_0=-\frac{d\cd}{dt}.
\end{equation}

\p On the other hand, maintaining the conservation of momentum supplies strong constraints, as before. Using the field equations in each branch, together with \eqref{junctionsewing}, leads to
\begin{eqnarray}\label{Mdot}\dot\cm&&=\sum_{i=1}^N\epsilon_i\left[\frac{c_i}{2}u^{(i)}_tu^{(i)}_t+\frac{1}{2c_i}\left(\sum_{j,k=1}^NA_{ij}A_{ik}u^{(j)}_tu^{(k)}_t\right.\right. \nonumber\\
&&\left.\left. \phantom{mmm} -2\sum_{j=1}^NA_{ij}u^{(j)}_t\frac{\partial\cd\phantom{^{(i)}}}{\partial u^{(i)}}+\frac{\partial\cd\phantom{^{(i)}}}{\partial u^{(i)}}\frac{\partial\cd\phantom{^{(i)}}}{\partial u^{(i)}}\right)-c_iU^{(i)}\right],
\end{eqnarray}
where all quantities are evaluated at the junction. Insisting the terms quadratic in field time-derivatives cancel requires the real antisymmetric matrix $A$ to satisfy a further constraint, which, in matrix form, is:
\begin{equation}\label{Aequation}
\epsilon c=A\epsilon c^{-1}A, \ {\rm or}\ \ (\epsilon c^{-1}A)^2=1, \ \ A=-A^T.
\end{equation}
Note, since the determinant of an odd-dimensional antisymmetric matrix is zero, the constraint \eqref{Aequation} cannot be satisfied if the junction is joining an odd number of branches. Thus, this scenario cannot, for example, work when there are three branches. On the other hand, the constraints might be satisfied if the number of branches is even.

\p For $N=2$, the defect introduced earlier, \eqref{Aequation} is satisfied by
\begin{equation}\label{A2expression}
\epsilon_1=-\epsilon_2,\ \ A=\left(\begin{array}{cc}
                                                                                                 0 & \sqrt{c_1 c_2}a \\
                                                                                                 -\sqrt{c_1 c_2}a & 0 \\
                                                                                               \end{array}
                                                                                             \right),\ \ a^2=1.\end{equation}

  \p  For $N=4$, the constraint \eqref{Aequation} can be solved and the solution depends on two free parameters and one discrete parameter. It can be written conveniently in the form:
\begin{equation}\label{A4expression}
A=\left( \begin {array}{cccc} 0&{ \epsilon_1} { \epsilon_2} \sqrt {{ c_1} {
 c_2}}\,a&{ \epsilon_1} { \epsilon_3} \sqrt {{ c_1} { c_3}}\,b&{ \epsilon_1} {
 \epsilon_4} \sqrt {{ c_1} { c_4}}\,c\\ \noalign{\medskip}-{ \epsilon_1} {
 \epsilon_2} \sqrt {{ c_1} { c_2}}\,a&0&\tau\sqrt {{ c_2} { c_3}}\,c&-\tau
\sqrt {{ c_2} { c_4}}\,b\\ \noalign{\medskip}-{ \epsilon_1} { \epsilon_3}
\sqrt {{ c_1} { c_3}}\,b&-\tau\sqrt {{ c_2} { c_3}}\,c&0&\tau\sqrt {{
 c_3} { c_4}}\,a\\ \noalign{\medskip}-{ \epsilon_1} { \epsilon_4} \sqrt {{
 c_1} { c_4}}\,c&\tau\sqrt {{ c_2} { c_4}}\,b&-\tau\sqrt {{ c_3} {
 c_4}}\,a&0\end {array} \right),
\end{equation}
where
$\tau$ is an arbitrary sign ($\tau^2=1$), $\epsilon_1\epsilon_2\epsilon_3\epsilon_4=1$, and $a,b,c$ are further constrained by the quadratic relation
\begin{equation}\label{abcconstraint}
\epsilon_1+\epsilon_2 \, a^2 +\epsilon_3\, b^2+\epsilon_4\, c^2=0.
\end{equation}
Since the product of the $\epsilon$'s is $+1$, and $a,b,c$ are real, the only possibility is that two $\epsilon$'s are positive and the other two are negative. Thus the constraint \eqref{abcconstraint} is a hyperbolic quadratic form reducing the number of free parameters in the matrix $A$ from three to two. Alternative expressions for matrices of this type will be given in section \ref{Free fields at a type I junction}.

\p Requiring the terms linear in the field time-derivatives in \eqref{Mdot} to be a total time derivative requires in turn:
\begin{equation}\label{DPequation}
\sum_{i=1}^N{\epsilon_i}{c_i^{-1}}\,\frac{\partial\cd\phantom{^{(i)}}}{\partial u^{(i)}}A_{ij}=\frac{\partial \cp\phantom{^{(j)}}}{\partial u^{(j)}},
\end{equation}
and removing the term that contains no time-derivatives of the fields requires
\begin{equation}\label{Dconstraint}
\sum_{i=1}^N\left(\frac{1}{2}\epsilon_i{c_i^{-1}}\,\frac{\partial\cd\phantom{^{(i)}}}{\partial u^{(i)}}\frac{\partial\cd\phantom{^{(i)}}}{\partial u^{(i)}}-\epsilon_ic_iU^{(i)}\right)=0.
\end{equation}
Equations \eqref{DPequation} and \eqref{Dconstraint} are generalisations of the type I defect equations \eqref{typeIconditions}.

\p As a consequence of \eqref{DPequation} and the antisymmetry of $A$, it follows that
\begin{equation}\label{Pcondition} \sum_{i=1}^N{\epsilon_i}{c_i^{-1}}\,\frac{\partial^2 \cp}{\partial u^{(i)}\partial u^{(i)}}=0.
\end{equation}
As a further consequence of \eqref{DPequation} and \eqref{Aequation}
\begin{equation}\label{PDequation}
\frac{\partial \cd\phantom{^{(j)}}}{\partial u^{(i)}}=\sum_{j=1}^N \frac{\partial \cp\phantom{^{(j)}}}{\partial u^{(j)}}{\epsilon_j}{c_j^{-1}} A_{ji},
\end{equation}
and hence,
\begin{equation}\label{Dcondition}
\sum_{i=1}^N{\epsilon_i}{c_i^{-1}}\,\frac{\partial^2 \cd}{\partial u^{(i)}\partial u^{(i)}}=0.
\end{equation}
It is tempting to try to write $\cd$ and $\cp$ as sums of exponentials of the form $e^{\alpha\cdot u}$, where
$\alpha\cdot u\equiv\sum_i\alpha_iu^{(i)}$, for suitably chosen vectors $\alpha$. However, in view of the relationships between $\cp$, and $\cd$ expressed by equations \eqref{PDequation} and \eqref{DPequation}, it is clear that $\alpha$ must be an eigenvector of the matrix $A\epsilon c^{-1}$. Because of the  relations \eqref{Aequation} the matrix $A\epsilon c^{-1}$ has two eigenvalues $\pm 1$, each with a two-dimensional eigenspace. Moreover, a pair of vectors $\alpha,\beta$ in either one of the eigenspaces have the property that $\alpha^T\epsilon c^{-1}\beta=0$, which, in turn, automatically implies equations \eqref{Pcondition} and \eqref{Dcondition}.

\p However, it does not seem generally possible to select a suitable set of eigenvectors in order to satisfy \eqref{Dconstraint} while allowing nonlinear integrable wave equations (such as sine-Gordon)  on all four legs meeting at the junction. A closer look into this possibility will be provided in section \ref{Non-linearity at a type I junction}. However, exceptions to this are the special cases where the four-point junction really consists of pairs of legs with one incoming and one outgoing leg in each pair. In essence, this happens when $A$ corresponds to a direct sum of $N=2$ cases. Even with a fixed set of epsilons (for example $\epsilon_1=\epsilon_3=-1=-\epsilon_2=-\epsilon_4$), there are two possibilities since the first branch could be paired with the second ($b=c=0$ in A) or fourth ($a=b=0$ in A) and the third branch then paired, respectively, with the fourth or second branches. In that sense, a junction could behave as a switch. The number of branches meeting at this type of junction is always even so this idea generalises allowing branches to be paired in a variety of ways by choosing the parameters in $A$ suitably.

\p
An interesting feature is the possibility of moving solitons around on a two-dimensional lattice network (using $N=4$ junctions), or on a three-dimensional lattice (using $N=6$ junctions).

\subsection{Free fields at a type I junction}
\label{Free fields at a type I junction}

\p In order to give a few more details it is useful to consider the simplest (non-conformal) situation where the network supports a collection of massive free fields. Then it is possible to go further with the analysis.

 \p The contributions to energy and momentum at the defect are given by quadratic expressions of the form
\begin{equation}\label{freedandp}
\cd=\frac{1}{2}\sum_{i,j=1}^Nd_{ij}u^{(i)}u^{(j)},\ \ \cp=\frac{1}{2}\sum_{i,j=1}^Np_{ij}u^{(i)}u^{(j)},
\end{equation}
where the two expressions are linked by \eqref{PDequation} and constrained by \eqref{Pcondition} and \eqref{Dcondition}. Hence, in matrix notation
$$p=d\epsilon c^{-1}A, \ \ {\rm tr}(\epsilon c^{-1}d)=0={\rm tr}(\epsilon c^{-1}p).$$
In fact, the latter two conditions are a consequence of the former using the facts that both $p$ and $d$ are symmetric matrices while $A$ is antisymmetric. For example,
$${\rm tr}(\epsilon c^{-1}p)={\rm tr}(\epsilon c^{-1}d\epsilon c^{-1}A)={\rm tr}(A^{T} c^{-1}\epsilon d^{T} c^{-1}\epsilon)=-{\rm tr}(\epsilon c^{-1}d\epsilon c^{-1}A)=-{\rm tr}(\epsilon c^{-1}p),$$
which implies ${\rm tr}(\epsilon c^{-1}p)=0$. Note also that because $p$ is symmetric
\begin{equation}\label{dAequation}
d\epsilon c^{-1}A+Ac^{-1}\epsilon d=0 \Rightarrow (\epsilon c^{-1}d)(\epsilon c^{-1}A)+(\epsilon c^{-1}A)(\epsilon c^{-1}d)=0.
\end{equation}
On the other hand, the constraint equation \eqref{Dconstraint} states in this case that
\begin{equation}\label{dequation}
d\epsilon c^{-1}d=\epsilon c m^2\ {\rm or}\ (\epsilon c^{-1}d)^2=m^2,
\end{equation}
where $m^2$ is a diagonal matrix whose entries are the free field mass parameters $m_i^2,\ i=1,\dots,N$. As a consequence of \eqref{dAequation} and \eqref{dequation}, the mass matrix $m^2$ must commute with $\epsilon c^{-1}A$. For example, when $N=4$, if any two of the parameters $a,b,c$ in the expression  \eqref{A4expression}  for the matrix $A$  are non-zero, this requires that $m^2$ be proportional to the identity matrix, meaning the four mass parameters are equal.

\p For $N=2$, given the expression for $A$ in \eqref{A2expression}, as already remarked, it is necessary that $m_1=m_2\equiv m$. Then, the symmetric matrix $d$ satisfying the equations \eqref{dequation} is conveniently parametrised by introducing a parameter $\eta$ so that
$$d=m\left(\begin{array}{cc} c_1\cosh\eta & \sqrt{c_1 c_2}\sinh\eta \\   \sqrt{c_1 c_2}\sinh\eta & c_2\cosh\eta \\  \end{array}  \right).$$
Then, on setting $\sigma = \exp(\eta)$ and $u^{(1)}\equiv u,\ u^{(2)}\equiv v$, the defect potential \eqref{freedandp} becomes
$${\cal D}=\frac{m\sigma}{4}\left(\sqrt{c_1}u+\sqrt{c_2}v\right)^2 + \frac{m}{4\sigma}\left(\sqrt{c_1}u-\sqrt{c_2}v\right)^2.$$
Note, up to a constant this is the quadratic part of  \eqref{DIsineGordon} and there is one free parameter.
\vspace{.5cm}

\p For $N=4$, the constraints represented by the two conditions \eqref{dAequation} and \eqref{dequation} are complicated. However, by redefining the fields to include a factor $\sqrt{c_i}$ and scaling the matrix $d$ by the common mass, the equations to be satisfied are
\begin{equation}\label{simplifiedAdequations}
(\epsilon A)^2=1, \quad (\epsilon d)^2=m^2, \quad (\epsilon d)(\epsilon A)+(\epsilon A)(\epsilon d)=0.
\end{equation}
In view of these relationships, it is convenient to set $\alpha=\epsilon A$ and $m\delta=\epsilon d$ so that,
\begin{equation}\label{alternativedAequations}
\alpha^{T}=-\epsilon\alpha\epsilon,\ \ \delta^{T}=\epsilon\delta\epsilon, \ \ \alpha^2=1,\ \ \delta^2=1, \ \ \alpha\delta+\delta\alpha=0.
\end{equation}
Then it is helpful to make use of the elements of a real Clifford algebra generated by
\begin{equation}\label{clifford1}
\gamma_1=i\sigma_2\otimes\sigma_1,\ \ \gamma_2=\sigma_1\otimes 1,\ \ \gamma_3=i\sigma_2\otimes\sigma_3,\ \ \gamma_4=\sigma_3\otimes 1,
\end{equation}
where $\sigma_a, \ a=1,2,3$ are the Pauli sigma matrices. With this choice
$$\{\gamma_a,\gamma_b\}=\epsilon_{ab},$$
where the matrix $\epsilon$ defined above plays the role of a metric, with $\epsilon={\rm diag}(-1,+1,-1,+1)=-\gamma_4$. Note also that $\gamma_2,\gamma_4$ are symmetric while $\gamma_1,\gamma_3$ are skew-symmetric. Then, a solution is constructed by assembling $\alpha$ and $\delta$ out of linear combinations of elements of the Clifford algebra while respecting \eqref{alternativedAequations}. For example,
\begin{equation}\label{parameterconstraints}
\alpha=a\gamma_2 +c\gamma_1\gamma_2\gamma_3\gamma_4,\ \ \delta=(q\gamma_1+r\gamma_3+s\gamma_4),\ \ a^2+c^2=1,\ \ s^2-q^2-r^2=1.
\end{equation}
Explicitly, this choice is:
\begin{equation}\label{simpleexample}
A=\epsilon\alpha=\left(
    \begin{array}{cccc}
      0 & -a & 0 & -c \\
      a & 0 & -c & 0 \\
      0 & c & 0 & -a \\
      c & 0 & a & 0 \\
    \end{array}
  \right),\ d=m\epsilon\delta=
  m\left(
    \begin{array}{cccc}
       -s& -r & 0 & -q \\
      -r & -s & -q & 0 \\
      0 & -q &-s  & r \\
      -q & 0 & r & -s \\
    \end{array}
  \right).
  \end{equation}
 This solution has three free parameters after taking the two constraints into account. \vspace{0.5cm}

 \p More generally, as noted previously ($\tau=\pm 1$ in \eqref{A4expression}),
 there are two possibilities for the matrix $\alpha$. Thus,
 $$\alpha_1=a\gamma_2 +b \gamma_1\gamma_3\gamma_4 +c\gamma_1\gamma_2\gamma_3\gamma_4,\  \ a^2-b^2+c^2=1;$$
 {\rm or,}$$ \alpha_2=a\gamma_1\gamma_3 +b \gamma_1\gamma_4 + c \gamma_3\gamma_4,\ \  b^2-a^2+c^2=1.$$
 For each of these there is a corresponding matrix $\delta$ depending on several free parameters. To see this, it is convenient to write
 \begin{equation}\label{delta}
 \delta=(u+v\gamma_2+w\gamma_1\gamma_2\gamma_3\gamma_4)(q\gamma_1+r\gamma_3+s\gamma_4),\ \ (u^2-v^2-w^2)(q^2+r^2-s^2)=-1.
 \end{equation}
Then, the second and fourth equations in the set \eqref{alternativedAequations} are automatic by construction while the fifth requires a single additional constraint. Choosing $\alpha_1$ or $\alpha_2$, the additional constraint is either
$$ aw-bu-cv=0\quad  {\rm or} \quad as-br+cq=0,$$
respectively. The special case above, given by the expression\eqref{simpleexample}), corresponds to $u=1,\ v=w=0$ in $\delta$, and $b=0$ in $\alpha_1$.

 \p It is now feasible to calculate a transmission matrix corresponding to plane waves on the four legs taking input data on legs $1,3$ and output data on legs $2,4$, for example, related by using the sewing conditions implied by \eqref{junctionsewing}. Written in the present context \eqref{junctionsewing} becomes:
 $$ c_i\sqrt{c_i}u_x^{(i)}=\sum_j\left(\alpha_{ij}\sqrt{c_j}u_t^{(j)}-m\delta_{ij}\sqrt{c_j}u^{(j)}\right)$$
 and the plane waves  (anticipating that there is no reflection on legs 1,3) are represented by
 $$u^{(i)}=u_0^{(i)}e^{i(k_i x-\omega t)},\ \ \omega^2=c_i^2k_i^2+m^2\equiv \kappa^2+m^2,\ \ i=1,2,3,4.$$
 Thus, the sewing condition can be rewritten as a matrix equation
 $$(i\kappa  +i\omega\alpha+m\delta)\sqrt{c}u_0\equiv M\sqrt{c}u_0 =0, \ \ u_0^T=(u_0^{(1)},u_0^{(2)},u_0^{(3)},u_0^{(4)}).$$
 It is straightforward to check that $M$ is singular (in fact $M^2-2i\kappa M=0$, which follows directly from the properties of the matrices $\alpha$ and $\delta$), and then, for the simplest of the examples above \eqref{simpleexample}:
 \begin{equation}
 \left(
   \begin{array}{c}
     \sqrt{c_2}u_0^{(2)} \\
      \sqrt{c_4}u_0^{(4)} \\
   \end{array}
 \right)=\frac{1}{ms-i\kappa}\left(
                               \begin{array}{cc}
                                 ia\omega - mr & -ic\omega -mq \\
                                 ic\omega-mq & ia\omega+mr \\
                               \end{array}
                             \right)
 \left(
   \begin{array}{c}
     \sqrt{c_1}u_0^{(1)} \\
    \sqrt{c_3} u_0^{(3)} \\
   \end{array}
 \right)\equiv T \left(
   \begin{array}{c}
      \sqrt{c_1}u_0^{(1)} \\
     \sqrt{c_3} u_0^{(3)} \\
   \end{array}
 \right).
 \end{equation}
Using the constraints and the relationship between $\kappa$ and $\omega$ it is straightforward to check that $TT^\dagger=I=T^\dagger T$. Since the eigenvalues of $M$ are $0,2i\kappa$, each with multiplicity two, there can be no reflection along the branches 1 and 3. If there is no incoming wave along branch 3 then the wave on branch 1 is split typically into a pair of waves along branches 2 and 4. Clearly, in that case the ratio of the amplitudes of the outgoing waves is given by:
$$\frac{\left| c_2 u_0^{(2)}\right|^2}{\left| c_4 u_0^{(4)}\right|^2}=\frac{a^2\omega^2+m^2 r^2}{c^2\omega^2+m^2 q^2}\ .$$
At the pole $\kappa=-ims$, the frequency would be given by $\omega^2=m^2(1-s^2)$. However, the constraint \eqref{parameterconstraints} then implies $\omega^2\le 0$, meaning there can be no bound state solution.
\vspace{0.5cm}

\p It is intriguing that a Clifford algebra seems to provide a natural context to capture the details of this kind of defect.

\subsection{Non-linearity at a type I junction}
\label{Non-linearity at a type I junction}

In a situation where the fields are conformal (either massless free or Liouville), there are opportunities for defects that join free to free, Liouville to Liouville, or free to Liouville \cite{bcz2003}. This means that since it is possible to have a junction with free massive (or massless) fields, it must be possible to have a junction with an even number of branches supporting Liouville fields. This is because the massless fields on any branch can be linked by a defect to a Liouville field at a point on a specific branch very close to a junction.

\p To explore this situation further, the case $N=4$ will be considered when the fields on each branch are either free massless fields or satisfy the Liouville equation of motion. As in the previous section, the matrix $\epsilon$ is chosen to be $\epsilon=\mbox{diag}(-1,+1,-1,+1),$ and the matrix $A$ is:
\begin{equation*}
A=\left(
  \begin{array}{ccccc}
   0& -a & b & -c\\
    a & 0 & -c  & b \\
    -b & c&0& -a \\
    c & -b& a & 0 \\
  \end{array}
\right),\quad a^2-b^2+c^2=1.
\end{equation*}
For convenience, the fields will be redefined to include a factor $\sqrt{c_i}$ and the Liouville potential is taken to be:
$$U^{(i)}=\frac{1}{2}c_i^{-1}\,l_i^2\,e^{2\beta_i u^{(i)}}.$$

\p General expressions for the potentials ${\cal D}$ and ${\cal P}$ are:
$${\cal D}=\sum_{k}x_k\, e^{\alpha_k\cdot u},\,\quad {\cal P}=\sum_{k}y_k\, e^{\alpha_k\cdot u},\,\quad u^T=(u^{(1)},u^{(2)},u^{(3)},u^{(4)}),$$
where $x_k,$ $y_k$ are constants and $\alpha_k$ are unspecified 4-components vector defined at the junction. All these new elements will be constraints by expressions \eqref{Dconstraint}-\eqref{Dcondition}.
Start with condition \eqref{PDequation}, which implies
\begin{equation}\label{LiouvillePDequation}
H^T\,\alpha_k=z_k\,\alpha_k,\quad H=\epsilon\, A,\quad z_k=\frac{y_k}{x_k}.
\end{equation}
Note that since $H^2=I,$ the eigenvalues $z_k$ are $\pm 1.$ Then, condition \eqref{Dcondition} translates into
\begin{equation}\label{LiouvilleDcondition}
\alpha_k\cdot\,\epsilon \, \alpha_k=0.
\end{equation}
On the other hand, the quadratic relation \eqref{Dconstraint} becomes
\begin{equation}\label{LiouvilleDconstraint}
\sum_{k,l}x_k\,x_l\,e^{(\alpha_k+\alpha_l)\cdot u}\,\alpha_l\cdot\,\epsilon \, \alpha_k=2\,( -c_1U^{(1)}+
 c_2U^{(2)}- c_3U^{(3)}+ c_4U^{(4)}),
\end{equation}
where the exponentials on the left hand side, whose coefficients are different form zero, must reproduce the Liouville potentials.
Because $\alpha_k$ are eigenvectors of $H$ with eigenvalues $\pm 1,$ the condition \eqref{LiouvilleDcondition} is always satisfied and the expression $\alpha_l\cdot\,\epsilon \, \alpha_k$ are always zero if $\alpha_k$ and $\alpha_l$ lay in the same two-dimensional eigenspace. Then, in order to reproduce the Liouville potentials, the sum of two eigenvectors belonging to different eigenspeces  must be equal to a vector with only one entry different from zero. In principle, that seems possible. By demanding, for instance, the field $u^{(1)}$ to be Liouville and the remaining fields free massless, and choosing
$$\alpha_1^T=\beta_1\,(1,a,-b,c),\quad\alpha_2^T=\beta_1\,(1,-a,b,-c),\quad\rightarrow\quad  (\alpha_1+\alpha_2)^T=(2\beta_1,0,0,0),$$
then
$$
\alpha_1\cdot\,\epsilon \, \alpha_2=-2\beta_1^2,\quad x_1x_2=\frac{l_1^2}{4\,\beta_1^2},
$$
and
 the potentials ${\cal D}$ and ${\cal P}$ are:
\begin{eqnarray*}
{\cal D}&=&\frac{l_1}{2\beta_1}\,e^{\beta\,u^{(1)}}\,\left(\sigma_1\,e^{\beta_1(au^{(2)}-bu^{(3)}+cu^{(4)})}+\frac{1}{\sigma}\,e^{-\beta_1(au^{(2)}-bu^{(3)}+cu^{(4)})}\right),\\
{\cal P}&=&\frac{l_1}{2\beta_1}\,e^{\beta\,u^{(1)}}\,\left(\sigma_1\,e^{\beta_1(au^{(2)}-bu^{(3)}+cu^{(4)})}-\frac{1}{\sigma}\,e^{-\beta_1(au^{(2)}-bu^{(3)}+cu^{(4)})}\right).
\end{eqnarray*}
On the other hand, by demanding that more than one field is Liouville, additional constraints on the entries of the matrix $H$ emerge. With Liouville fields in two of the branches, one of the constants $a,$ $b,$ $c$ is forced to be zero. For instance, setting $a=0,$ the eigenvectors can be
$$\alpha_1^T=\beta_4\,(c,-b,0,1),\quad \alpha_2^T=\beta_3\,(b,-c,1,0),\quad
\alpha_3^T=\beta_4\,(-c,b,0,1),\quad \alpha_4^T=\beta_3\,(-b,c,1,0).$$
Then
$$\alpha_1\cdot\,\epsilon \, \alpha_3=2\beta_4^2,\quad \alpha_2\cdot\,\epsilon \, \alpha_4=-2\beta_3^2,\quad
x_1x_3=\frac{l_4^4}{4\,\beta_4^2},\quad x_2x_4=\frac{l_3^4}{4\,\beta_3^2},$$
$u^{(3)}$ and $u^{(4)}$ are Liouville and  $u^{(1)}$ and $u^{(2)}$ are free massless. The potentials ${\cal D}$ and ${\cal P}$ are:
\begin{eqnarray*}
{\cal D}&=&\frac{l_3}{2\beta_3}\,e^{\beta_3\,u^{(3)}}\,\left(\sigma_3\,e^{\beta_3(cu^{(1)}-cu^{(2)})}+\frac{1}{\sigma_3}\,e^{-\beta_3(cu^{(1)}-cu^{(2)})}\right)\\
&&+\frac{l_4}{2\beta_4}\,e^{\beta_4\,u^{(4)}}\,\left(\sigma_4\,e^{\beta_4(cu^{(1)}-bu^{(2)})}+\frac{1}{\sigma_4}\,e^{-\beta_4(cu^{(1)}-bu^{(2)})}\right),\\
{\cal P}&=&\frac{l_3}{2\beta_3}\,e^{\beta_3\,u^{(3)}}\,\left(\sigma_3\,e^{\beta_3(cu^{(1)}-cu^{(2)})}-\frac{1}{\sigma_3}\,e^{-\beta_3(cu^{(1)}-cu^{(2)})}\right)\\
&&+\frac{l_4}{2\beta_4}\,e^{\beta_4\,u^{(4)}}\,\left(\sigma_4\,e^{\beta_4(cu^{(1)}-bu^{(2)})}-\frac{1}{\sigma_4}\,e^{-\beta_4(cu^{(1)}-bu^{(2)})}\right).
\end{eqnarray*}
Note that in all these examples, all four fields interact at the junction, and such interactions are controlled by the entries of the $H(=\epsilon\, A)$ matrix.
Finally, if all fields are required to be Liouville, then a second constant in the matrix $H$ has to be set to zero and the remaining one is forced to be $\pm 1.$
In essence, the system splits into the sum of two independent sets with two branches each. As noted earlier in section \ref{Junctions on a graph}, the latter is clearly the only possibility if the Liouville fields are replaced by sine-Gordon fields. Hence, in either case, the system reduces to two sets of type I defects.

\subsection{Linear Schr\"odinger equation}

\p A similar analysis can be carried out for a linear Schr\"odinger equation but there are some differences. On each leg at a junction, the appropriate equation is
$$iu^{(k)}_t=-\lambda_k u^{(k)}_{xx},\ \ \lambda_k>0,\ \ k=1,2,\dots, N.$$
It is found that all $\lambda_k$ are forced to be the same, hence it is convenient  to  scale variables and take $\lambda_k=1, \ k=1,2,\dots,N$ from the start. Then, the linear sewing conditions, at $x=x_0$,
\begin{equation}\label{Ssewingconditions}
u^{(k)}_x=\sum_1^N(iA_{kl}u^{(l)}_t+B_{kl}u^{(l)}),
\end{equation}
are consistent with (modified) conservation laws for the quantities
\begin{eqnarray}
&&{\cal N}=\sum_k\epsilon_k\int^{\infty}_{x_0}\bar u^{(k)}u^{(k)} dx -\sum_{kl}\left[\bar u^{(k)}\epsilon_k A_{kl}u^{(l)}\right]_{x_0},\\
&&{\cal E}=\sum_k\epsilon_k\int^{\infty}_{x_0}\bar u_x^{(k)}u^{(k)}_x dx +\sum_{kl}\left[\bar u^{(k)}\epsilon_k B_{kl}u^{(l)}\right]_{x_0},\\
 &&{\cal P}=\sum_k\epsilon_k\int^{\infty}_{x_0}i(\bar u^{(k)} u^{(k)}_x-\bar u^{(k)}_x u^{(k)})dx -\sum_{kl}\left[i\bar u^{(k)}\epsilon_k (AB-BA)_{kl}u^{(l)}\right]_{x_0} \\
 \end{eqnarray}
provided the matrices $A,\ B$ satisfy
\begin{equation}\label{ABconstraints}
A^\dagger=\epsilon A\epsilon,\ \ B^\dagger=\epsilon B\epsilon, \ \ A^2=0, \ \ B^2=0,\ \ AB+BA=-1.
\end{equation}
 The case $N=2$ was considered previously  as a special case of a defect inserted in the nonlinear Schr\"odinger equation \cite{cz2005}. To be explicit, as a straightforward example for $N=2$, it is convenient to take $\epsilon=-\sigma_3$, and
 $$A=\frac{1}{2\alpha}(i\sigma_2 -\sigma_3),\ \ B=\frac{\alpha}{2}(i\sigma_2+\sigma_3),$$
 where $\alpha$ is a real free parameter. More generally, the constraints on the matrices $A$ and $B$ can be solved by setting
 \begin{eqnarray}\label{N=2ABconstraints}
& A=i\alpha_1\sigma_1+i\alpha_2\sigma_2+\alpha_3\sigma_3, \quad B=i\beta_1\sigma_1+i\beta_2\sigma_2+\beta_3\sigma_3,\nonumber\\
 & \alpha_3^2=\alpha_1^2+\alpha_2^2,\quad  \beta_3^2=\beta_1^2+\beta_2^2, \quad \alpha_1\beta_1+\alpha_2\beta_2-\alpha_3\beta_3=\frac{1}{2},
 \end{eqnarray}
 where $\alpha_i,\beta_i,\ i=1,2,3$ are real parameters. This means the general solution to the constraints for $N=2$ has three real free parameters. To make these explicit, the constraints \eqref{N=2ABconstraints} can be solved conveniently by setting
 \begin{eqnarray}&\alpha_1=\alpha_3\sin\theta, \ \alpha_2=\alpha_3\cos\theta, \ \beta_1=\beta_3\sin\phi,\ \beta_2=\beta_3\cos\phi,\nonumber\\  &\alpha_3=\frac{1}{2\alpha}\left(\sin\left(\frac{\theta-\phi}{2}\right)\right)^{-1}, \ \beta_3=-\frac{\alpha}{2}\left(\sin\left(\frac{\theta-\phi}{2}\right)\right)^{-1},
 \end{eqnarray}
 where $\theta-\phi$ is not zero or a multiple of $2\pi$. The simpler case above, with just one parameter, corresponds to $\theta=\pi,\ \phi=0$. With this parametrisation, the transmission factor for a monochromatic wave is:
 \begin{equation}
 T(k,\alpha,\theta,\phi)=-e^{-i\theta}\left(\frac{k-\alpha e^{i(\theta-\phi)/2}}{k-\alpha e^{-i(\theta-\phi)/2}}\right).
 \end{equation}

 \p If $N=3$ then, as before, a direct calculation reveals there is no solution to the constraints.
On the other hand, for the case $N=4$, it is useful to take advantage of the Clifford algebra defined previously in \eqref{clifford1}, with $\epsilon\equiv\gamma_4$, and put
 $$A=a_1\gamma_1+ia_2\gamma_2+a_3\gamma_3+a_4\gamma_4+i a_5 \gamma_1\gamma_2\gamma_3\gamma_4,\ \ B=b_1\gamma_1+ib_2\gamma_2+b_3\gamma_3+b_4\gamma_4+i b_5 \gamma_1\gamma_2\gamma_3\gamma_4,$$
 where all coefficients are real.
 Then the constraints \eqref{ABconstraints} are satisfied provided
 \begin{eqnarray}
 &a_1^2+a_2^2+a_3^2-a_4^2 +a_5^2=0, \ \  b_1^2+b_2^2+b_3^2-b_4^2 +b_5^2=0,\nonumber \\
 & a_1b_1+a_2b_2+a_3b_3-a_4b_4+a_5b_5=\frac{1}{2}.
 \end{eqnarray}
Here, there are ten real parameters with just three real constraints that need to be satisfied. Again, it is noteworthy not only that a Clifford algebra provides a natural setting but also the constraints are a pair of light cones in a five-dimensional Minkowski space with signature $(1,1,1,-1,1)$ (with the labelling used above).

\section{Type II junctions in a network}

\p In this section the feasibility of a type II junction is explored, which means additional degrees of freedom are allowed at the junction.

\subsection{The setting}
\label{Type II junctions on a graph}

\p Conservation of energy and momentum is examined afresh in Appendix A and an alternative set of conditions are found. For this section, the starting point is provided by the sewing conditions \eqref{JunctionNewConditionsAppendix1}-\eqref{JunctionNewConditionsAppendix3} with $A=\tilde{A}=B=0.$  In other words, \eqref{junctionsewing} is replaced by:
\begin{eqnarray}
&&c_i^2u_x^{(i)}=\epsilon_i\left(\sum_{j=1}^n\lambda_t^{(j)}-\frac{\partial {\cal D}}{\partial u^{(i)}}\right),\quad i=1,\dots,n,\label{JunctionNewConditions1}\\
&&\tilde{c}_k^2\tilde{u}_x^{(k)}=\tilde{\epsilon}_k\left(\sum_{j=1}^n\tilde{C}_{kj} \lambda_t^{(j)}-\frac{\partial {\cal D}}{\partial \tilde{u}^{(k)}}\right),\quad k=1,\dots,N-n,\label{JunctionNewConditions2}\\
&& u_t^{(i)}+\sum_{k=1}^{N-n}\tilde{u}_t^{(k)}\,\tilde{C}_{ki}=-\frac{\partial {\cal D}}{\partial \lambda^{(i)}},\quad i=1,\dots,n,\label{JunctionNewConditions3}
\end{eqnarray}
where $u^{(i)}$ are $n$ fields, $\tilde{u}^{(k)}$ are $ (N-n)$ fields and $\lambda^{(i)}$ are $n$ auxiliary fields defined only at the junction. Hence, $\tilde{C}$ is an $((N-n)\times n)$ matrix, which satisfies the following conditions
\begin{equation}\label{constraintTypeII_1}
{\epsilon}{c}^{-1}+\tilde{C}^T\,\tilde{\epsilon}{\tilde{c}^{-1}}\,\tilde{C}=0,
\end{equation}
and
\begin{equation}\label{constraintTypeII_3}
\tilde{\epsilon}\,\tilde{c}+\tilde{C}\,\epsilon\,c\,\tilde{C}^T=0.
\end{equation}
The conservation of the scaled momentum implies
\begin{equation}\label{XandY}
\dot{\cal M}={\cal X}+{\cal Y}=-\frac{d{\cal P}}{dt},
\end{equation}
where
\begin{eqnarray*}
{\cal X}&=&\sum_{i=1}^n\left(\frac{\epsilon_i}{2}{\,c_i}^{-1}\left(\frac{\partial {\cal D}}{\partial \,u^{(i)}}\right)^2+\frac{1}{2}\,\epsilon_i\,c_i\,\left(\frac{\partial {\cal D}}{\partial \,\lambda^{(i)}}\right)^2-\epsilon_i\,c_iU^{(i)}\right)\\
&&\phantom{m}+\sum_{k=1}^{N-n}\left(\frac{\tilde{\epsilon}_k}{2}\,{\tilde{c}_k}^{-1}\left(\frac{\partial {\cal D}}{\partial \,\tilde{u}^{(k)}}\right)^2
-\tilde{\epsilon}_k\,\tilde{c}_k\tilde{U}^{(k)}\right),
\end{eqnarray*}
and
\begin{equation*}
{\cal Y}=-\sum_{j=1}^{n}\sum_{k=1}^{N-n}\left(\lambda_t^{(j)}\left(\tilde{C}^T\,{\tilde{\epsilon}}{\tilde{c}}^{-1}\right)_{jk}\frac{\partial {\cal D}}{\partial \tilde{u}^{(k)}}+\lambda_t^{(j)}\left(\epsilon{c}^{-1}\right)_{jk}\,\frac{\partial {\cal D}}{\partial u^{(k)}}
-\tilde{u}_t^{(j)}\left(\tilde{C}\,\epsilon c \right)_{jk}\,
\frac{\partial {\cal D}}{\partial \lambda^{(k)}}\right).
\end{equation*}
In order to deal with the constraints imposed by the conservation of the scaled momentum, it is useful to perform a change of variables. Expression \eqref{JunctionNewConditions3} suggests setting
\begin{equation}\label{cofv_q}
q^{(i)}=u^{(i)}+\sum_{k=1}^{N-n}\tilde{u}^{(k)}\, \tilde{C}_{ki},\quad \mbox{with} \quad q^{(i)}_t=-\frac{\partial {\cal D}}{\partial \lambda^{(i)}},\quad i=1,\dots,n.
\end{equation}
In addition,
\begin{equation}\label{cofv_p}
p^{(i)}=\sum_{j=1}^{n}u^{(j)}\,F_{ji}+\sum_{k=1}^{N-n}\tilde{u}^{(k)}\,\tilde{F}_{ki},\quad i=1,\dots,n,
\end{equation}
where $F,$ $\tilde{F}$ are  matrices of dimensions $(n\times n)$  and $((N-n)\times n),$ respectively. In order to apply such a change of variables, it is necessary to assume that the matrices involved are invertible, which forces $n=N/2.$ Hence, the number of branches is even and the number of auxiliary fields is half the total number of branches meeting at the junction. Then, using the properties \eqref{constraintTypeII_1} and forcing the coefficient of the terms quadratic in the derivative of ${\cal D}$ with respect to the auxiliary fields to be zero, it is found that $\tilde{F}=-\tilde{C}F$ and \eqref{XandY} become
\begin{eqnarray}\label{typeIIQuadraticRelation}
&&\phantom{mm}\sum_{i=1}^n\left(\frac{\partial {\cal D }}{\partial q^{(i)}}\,\frac{\partial {\cal P}}{\partial \lambda^{(i)}}
-\frac{\partial {\cal P}}{\partial q^{(i)}}\,\frac{\partial {\cal D}}{\partial \lambda^{(i)}}-\epsilon_i\,c_i\,U^{(i)}
-\tilde{\epsilon}_i\,\tilde{c}_i\,\tilde{U}^{(i)}\right)=0,\\
&&\sum_{j=1}^n\left(2\,{\epsilon}{c}^{-1}\,F\right)_{ij}\,\frac{\partial {\cal D}}{\partial p^{(j)}}=\frac{\partial {\cal P}}{\partial \lambda^{(i)}},\quad
\sum_{j=1}^n\left(\frac{1}{2}\,F^{-1}\,\epsilon\,c\right)_{ij}\,\frac{\partial {\cal D}}{\partial \lambda^{(j)}}=\frac{\partial {\cal P}}{\partial p^{(i)}},\quad i=1,\dots,n.\nonumber\\\label{typeIILinearRelations}
\end{eqnarray}
Since the matrix $F$ is not yet determined, it is convenient to set
$$2\,{\epsilon}{c}^{-1}\,F=I,\quad \frac{1}{2}\,F^{-1}\,\epsilon\,c=I,$$
hence
$$F=\frac{1}{2}\epsilon\,c,\quad \tilde{F}=-\frac{1}{2}\tilde{C}\epsilon\,c,\quad p^{(i)}=\frac{1}{2}\left(u^{(i)}-\sum_{j=1}^{n}\tilde{u}^{(j)}\,\tilde{C}_{ji}\right)\,\epsilon_i\,c_i, \quad i=1,\dots,n.$$
The advantage of such a choice is that conditions \eqref{typeIILinearRelations} are satisfied by potentials of the following kind
\begin{equation}\label{FG}
{\cal D}={\cal F}(p+\lambda,q)+{\cal G}(p-\lambda,q),\quad {\cal P}={\cal F}(p+\lambda,q)-{\cal G}(p-\lambda,q).
\end{equation}

\p It must be remembered that the matrix $\tilde{C}$ must satisfy constraints \eqref{constraintTypeII_1} and \eqref{constraintTypeII_3}. By setting
\begin{equation}\label{CtildeC}
\tilde{C}=\sqrt{\tilde{c}}\,C\,(\sqrt{c})^{-1},
\end{equation}
the constraints become
$$C^T\,\tilde{\epsilon}\, C=-\epsilon,\quad C\epsilon\, C^T=-\tilde{\epsilon}.$$
For $N=3$ there are no solutions. For $N=4$, solutions are possible provided $n=2.$ Then all matrices are quadratic and  $|C|^2=|\epsilon||\tilde{\epsilon}|,$ since $N/2$ is even. Then $|\epsilon|=|\tilde{\epsilon}|$ and $C$ is invertible. If $C$ is invertible the contraints above are identical. In summary, for $N=4,$ $\epsilon_1\epsilon_2\tilde{\epsilon}_1\tilde{\epsilon}_2=1$ and
\begin{eqnarray}
C= \left(\begin{array}{cc}
a & \tau_2 c \\
c & \tau_1 a
\end{array}\right),\quad \tau_1^2=\tau_2^2=1,\quad |C|=\tau_1 a^2 -\tau_2 c^2= \tau_2\epsilon_1\tilde{\epsilon}_2,\quad \tau_1 \tau_2\epsilon_1\epsilon_2=-1.\label{C_solution}
\end{eqnarray}
Looking in more details at the possible values of $\epsilon_i$ and $\tilde{\epsilon}_i,$ the previous solution gives rise to the following cases
\begin{eqnarray}\label{C_cases}
&\epsilon_1\,\tilde{\epsilon}_2=1,\quad \tau_1\tau_2=1,\quad \epsilon_1\,\epsilon_2=-1,\quad \epsilon_2\,\tilde{\epsilon}_2=-1,&\quad a^2-c^2=1\nonumber\\
&\epsilon_1\,\tilde{\epsilon}_2=-1,\quad \tau_1\tau_2=1,\quad \epsilon_1\,\epsilon_2=-1,\quad \epsilon_2\,\tilde{\epsilon}_2=1,&\quad c^2-a^2=1,\nonumber\\
&\epsilon_1\,\tilde{\epsilon}_2=-1,\quad \tau_1\tau_2=-1,\quad \epsilon_1\,\epsilon_2=1,\quad \epsilon_2\,\tilde{\epsilon}_2=-1,&\quad c^2+a^2=1.
\end{eqnarray}
Notice that setting $c=0$ the following diagonal solution is obtained
\begin{equation}
C=\left(\begin{array}{cc}
a & 0 \\
0 & \tau\,a
\end{array}\right),\quad a^2=1,\quad \tau^2=1,\quad \epsilon_1\tilde{\epsilon}_1=-1.\label{C_solutionDiagonal}
\end{equation}
Because $C$ is invertible, it is possible to make use of the change of variables performed previously and operate in terms of fields $q^{(i)},$ $p^{(i)}$ and $\lambda^{(i)}$ with $i=1,2.$

\subsection{Free fields at a type II junction}

\p Consider the case when the fields on the branches are massive free fields. Following the observation \eqref{CtildeC}, it is useful to define a new set of $q,$ $p,$ $\lambda$ variable by performing a simple reascaling. The new set will be $q\,(\sqrt{c})/2,$ $p\,(\sqrt{c})^{-1},$ $\lambda\,(\sqrt{c})^{-1}.$ Then
\begin{equation}\label{New_q_and_p}
q^T=\frac{1}{2}\left(u\,\sqrt{c}\,I +\tilde{u}\,\sqrt{\tilde{c}}\, C\right)^T,\quad (\epsilon\,p)^T=\frac{1}{2}\left(u\,\sqrt{c}\, I-\tilde{u}\,\sqrt{\tilde{c}} \,C\right)^T,
\end{equation}
which implies
\begin{equation}\label{New_u_and_uTilde}
(u\,\sqrt{c})^T=(\epsilon\,p+q)^T,\quad
(\tilde{u}\,\sqrt{\tilde{c}})^T=(q\,C^{-1}-\epsilon\,p\,C^{-1})^T.
\end{equation}
It is worth stressing that  vectors $u,$ $\tilde{u},$ $q,$ $p,$ $\lambda$ are two dimensional and matrices $C,$ $\epsilon,$ $c,$ $\tilde{c}$ are $(2 \times 2).$
The rescaling does not change \eqref{typeIILinearRelations}, hence the potentials have the form \eqref{FG}. On the other hand,  \eqref{typeIIQuadraticRelation} reads
\begin{equation}\label{FGquadratic}
\sum_{i=1}^2\left(\frac{\partial {\cal D}}{\partial q^{(i)}}\,\frac{\partial {\cal P}}{\partial \lambda^{(i)}}
-\frac{\partial {\cal P}}{\partial q^{(i)}}\,\frac{\partial {\cal D}}{\partial \lambda^{(i)}}\right)=2\,\sum_{i=1}^2\,\left(\epsilon_i\,c_i\,U^{(i)}
+\tilde{\epsilon}_i\,\tilde{c}_i\,\tilde{U}^{(i)}\right).
\end{equation}
 The functions ${\cal F}$ and ${\cal G}$ in \eqref{FG} will have the form
\begin{eqnarray*}
{\cal F}&=&\frac{1}{2}(p+\lambda)^TF_1(p+\lambda)+\frac{1}{2}q^TF_2\,q+(p+\lambda)^TF_3\,q,\\
{\cal G}&=&\frac{1}{2}(p-\lambda)^TG_1(p-\lambda)+\frac{1}{2}q^TG_2\,q+(p-\lambda)^TG_3\,q,
\end{eqnarray*}
where $F_1,$ $F_2,$ $G_1,$ $G_2,$ are symmetric matrices
and the right hand side of \eqref{FGquadratic} becomes
$$u^T\,\epsilon \,M\,u+\tilde{u}^T\,\tilde{\epsilon}\,\tilde{M}\,\tilde{u} = q^T\,N_1\,q+p^T\,N_1\,p+2\,q^T\,N_2\,p, $$
with $M=$diag$(m_1^2,m_2^2),$ $\tilde{M}=$diag$(\tilde{m}_1^2,\tilde{m}_2^2)$ and
$$N_1=\epsilon \,M+\epsilon \,C^{-1}\,\tilde{M}\,\tilde{\epsilon}\,(C^{-1})^{T}\,\epsilon,\quad
\epsilon \,N_2=\epsilon \,M-\epsilon \,C^{-1}\,\tilde{M}\,\tilde{\epsilon}\,(C^{-1})^{T}\,\epsilon.$$
Because of the absence of auxiliary fields on the right hand side of \eqref{FGquadratic}, the matrix $N_1$ must be zero.  In the most general case, that is when both $a$ and $c$ are different from zero, all masses are forced to have the same value $m.$ On the other hand, the conditions on the masses are $\tilde{m}_1=m_1$ and $\tilde{m}_2=m_2$ if either $c$ or $a$ are zero. The matrix $N_2$ simplifies and becomes $N_2=$diag$(2\,m^2,2\,m^2),$ or $N_2=$diag$(2\,m_1^2,2\,m_2^2),$ respectively.
Finally, the additional constraints imposed by \eqref{FGquadratic} are:
\begin{eqnarray}
&&F_2\,G_3+G_2\,F_3=F_3\,G_1+(G_3\,F_1)^T=0,\nonumber\\
&&\phantom{mm} F_3\,G_3+F_1\,G_2=G_3\, F_3+G_1\,F_2,\nonumber\\
&& \phantom{mmmm}2\,(F_3\,G_3+F_1\,G_2)=N_2.
\end{eqnarray}
Solutions can be found in Appendix B. In this section the following simple solution
$$
F_1=G_1=\left(\begin{array}{cc}
0 & \alpha \\
\alpha & 0
\end{array}\right),\quad F_2=G_2=\left(\begin{array}{cc}
0 & m^2/\alpha \\
 m^2/\alpha & 0
\end{array}\right),\quad F_3=G_3=0,
$$
with $m=m_1=m_2,$ will be considered
since it suffices to show the presence of bound states at the junction. Choose $C=\epsilon=\mbox{diag}(1,1)=-\tilde{\epsilon},$ with $\tilde{u}^{(1)},$ $\tilde{u}^{(2)}$ the incoming plane waves. Then,  the transmission matrix at the junction, located at $x=0,$ is:
$$
 \left(
   \begin{array}{c}
     \sqrt{c_1}\,u_0^{(1)} \\
     \sqrt{c_2}\,u_0^{(2)} \\
   \end{array}
 \right)=\frac{1}{(i\kappa-\alpha)\,(i\kappa+\alpha)}\,\left(\begin{array}{cc}
(\kappa-\alpha)\,(\kappa+\alpha) &i\,2\kappa\,\alpha\\
i\,2\kappa\,\alpha & (\kappa-\alpha)\,(\kappa+\alpha)
\end{array}\right)
 \left(
   \begin{array}{c}
     \sqrt{\tilde{c}_1}\,\tilde{u}_0^{(1)} \\
     \sqrt{\tilde{c}_2}\,\tilde{u}_0^{(2)} \\
   \end{array}
 \right).
$$
A bound state for this solution is:
\begin{eqnarray*}
\tilde{u}^{(1)}=\tilde{u}_0^{(1)}e^{-i\omega t}\,e^{\alpha x/\tilde{c}_1},\quad x<0;\quad u^{(1)}=0,\quad x>0,\\
\tilde{u}^{(2)}=\tilde{u}_0^{(2)}e^{-i\omega t}\,e^{\alpha x/\tilde{c}_2},\quad x<0;\quad u^{(2)}=0,\quad x>0,
\end{eqnarray*}
with $\alpha^2<m^2.$  The details of the calculation have been relegated to Appendix B.

\section{Non-linearity at a type II junction}

\p As mentioned before, at the beginning of section 3.3, the type II junction provides more space for manoeuvre than a type I junction and, as a consequence, a Liouville field can be located on each of the four branches of a four-branch junction. This idea will be explored further in this section.

\p Consider the matrix $C$ \eqref{C_solution}. Without lost of generality, choose
$$C= \left(\begin{array}{cc}
a & c \\
c & -a
\end{array}\right),\quad c^2+ a^2=1,$$
corresponding to the last case in \eqref{C_cases} with $\tau_1=-1,$ $\tau_2=1,$ $\epsilon_1=\epsilon_2=1, $  $\tilde{\epsilon}_1=\tilde{\epsilon}_2=-1.$
The fields $\tilde{u}^{(1)},$ $\tilde{u}^{(2)}$ will be relabeled as $u^{(3)},$ $u^{(4)},$ respectively. Then, the Liouville potential for a field $u^{(i)}$ is taken to be:
$$U^{(i)}=\frac{l_i^2}{2c_i}\,\,e^{2\beta_i \sqrt{c_i}\,u^{(i)}}.$$
Then, using \eqref{New_u_and_uTilde} the right hand side of relation \eqref{FGquadratic}
can be rewritten in terms of the two dimensional vectors $p$ and $q$ as
\begin{equation}\label{LiouvilleRHS}
2\,\sum_{j=1}^2(\epsilon_i\,c_i\,U^{(i)}
+\tilde{\epsilon_i}\,\tilde{c_i}\,\tilde{U^{(i)}})\equiv \sum_{j=1}^4 \omega_i\,c_i\,U^{(i)}=\sum_{k=1}^4\,l_k^2\,\omega_k\,e^{2\alpha_k\cdot (p+\omega_k q)},
\end{equation}
with
\begin{eqnarray*}
\alpha_1^T&=&\beta_1\,(1,0), \quad \alpha_2^T=\beta_2\,(0,1),\quad \omega_1=\omega_2=1,\\
\alpha_3^T&=&\beta_3\,(-a,-c), \quad \alpha_4^T=\beta_4\,(-c,a),\quad \omega_3=\omega_4=-1.
\end{eqnarray*}
\p Suitable expressions for the functions ${\cal F}$ and ${\cal G}$ are:
$${\cal F}=\sum_{k=1}^4\,x_k\, e^{\alpha_k\cdot (p+\lambda)}\,e^{\gamma_k\cdot q},\quad {\cal G}=\sum_{l=1}^4\,y_l\, e^{\alpha_l\cdot (p-\lambda)}\,e^{\delta_l\cdot q},$$
where $x_i,$ $y_i$ are constants and $\gamma_i,$ $\delta_i$ are two dimensional vectors constrained by the relation \eqref{FGquadratic}. The left hand side of this relation reads
\begin{eqnarray}\label{LiouvilleLHS}
&& \sum_{i=1}^2\left(\frac{\partial {\cal D}}{\partial q^{(i)}}\,\frac{\partial {\cal P}}{\partial \lambda^{(i)}}
-\frac{\partial {\cal P}}{\partial q^{(i)}}\,\frac{\partial {\cal  D}}{\partial \lambda^{(i)}}\right)\nonumber\\&&\qquad =
2\,\sum_{k,l=1}^4\,\,x_k\,y_l\,(\gamma_k\cdot\alpha_l+\delta_l\cdot\alpha_k)\,e^{(\alpha_k+\alpha_l)\cdot p+(\alpha_k-\alpha_l)\cdot\lambda+(\gamma_k+\delta_l)\cdot q}.
\end{eqnarray}
Expressions \eqref{LiouvilleRHS} and \eqref{LiouvilleLHS} must coincide.
Clearly the terms in \eqref{LiouvilleLHS} with $k=l$ provide the terms in \eqref{LiouvilleRHS} if
\begin{equation}
(\gamma_k+\delta_k)=2\,\omega_k\,\alpha_k,\quad 4\,x_k\,y_k\,(\alpha_k\cdot\alpha_k)=l_k^2.
\end{equation}
This implies
$$\delta_k=(2\,\omega_k\,\alpha_k-\gamma_k),\quad x_k\,y_k=\frac{l_k^2}{4\,\beta_k^2},$$
since $(\alpha_k\cdot\alpha_k)=\beta_k^2.$
The remaining terms in \eqref{LiouvilleLHS} have all different exponentials. It follows that they have to be zero independently, which implies
\begin{equation}
(\gamma_k\cdot\alpha_l+\delta_l\cdot\alpha_k)=0\quad \rightarrow\,\quad \gamma_l\cdot\alpha_k-\gamma_k\cdot\alpha_l=2\,\omega_l\,(\alpha_k\cdot\alpha_l),\quad k\neq l.
\end{equation}
By interchanging the indices $k$ and $l$ in these relations, the following compatibility conditions are obtained
$$(\alpha_k\cdot\alpha_j)\,(\omega_k+\omega_l)=0.$$
In the present case they are satisfies, as it can be easily verified. Hence, the remaining relations are:
\begin{eqnarray}
&(\alpha_1\cdot\gamma_2-\alpha_2\cdot\gamma_1)=0,\quad
&(\alpha_3\cdot\gamma_4-\alpha_4\cdot\gamma_3)=0,\nonumber\\
&(\alpha_1\cdot\gamma_3-\alpha_3\cdot\gamma_1)=2\,a\,\beta_1\,\beta_3,\quad &(\alpha_1\cdot\gamma_4-\alpha_4\cdot\gamma_1)=2\,c\,\beta_1\,\beta_4,\nonumber\\
&(\alpha_2\cdot\gamma_3-\alpha_3\cdot\gamma_2)=2\,c\,\beta_2\,\beta_3,\quad &(\alpha_2\cdot\gamma_4-\alpha_4\cdot\gamma_1)=-2\,a\,\beta_2\,\beta_4.
\end{eqnarray}
These relations constrain the vectors $\gamma_i.$ Setting
$$\gamma_1^T=(a_1, a_2),\quad \gamma_2^T=(b_1, b_2),\quad \gamma_3^T=(k_1, k_2),\quad \gamma_4^T=(l_1, l_2),$$
the constraints are
\begin{eqnarray}\label{LiouvilleMixing}
\beta_1\,b_1&=&\beta_2\,a_2,\nonumber\\
\beta_1\,l_1&=&\beta_3\,(2\,\beta_1\,a-a\,a_1-c\,a_2),\nonumber\\
\beta_1\,k_1&=&\beta_4\,(2\,\beta_1\,c-c\,a_1+a\,a_2),\nonumber\\
\beta_2\,l_2&=&\beta_3\,(2\,\beta_2\,c-c\,b_2-a\,b_1),\nonumber\\
\beta_2\,k_2&=&-\beta_4\,(2\,\beta_2\,a-a\,b_2+c\,b_1).
\end{eqnarray}
A couple of examples can be provided. For instance, setting
$b_1=a_2=0,$ $a_2=\beta_1,$\ $b_1=\beta_2,$
then $\gamma_i=\delta_i,$ for $i=1,2,3,4,$ $\alpha_k=\gamma_k,$ for $k=1,2$ and $\alpha_l=-\gamma_l$ for $l=3,4.$ Then, the function ${\cal F}$ and ${\cal G}$ become
\begin{eqnarray*}
{\cal F}=\frac{1}{2}\sum_{k=1}^4\,\frac{l_k\,\sigma_k}{\beta_k}\, e^{\alpha_k\cdot (p+\lambda+\omega_k q)},\quad
{\cal G}=\frac{1}{2}\sum_{k=1}^4\,\frac{l_k}{\sigma_k\,\beta_k}\, e^{\alpha_k\cdot (p-\lambda+\omega_k q)},
\end{eqnarray*}
and the defect potential, in terms of the fields $u^{(i)},$ which have been redefined to include a factor $\sqrt{c_i,}$ is:
\begin{eqnarray}
{\cal D}={\cal F}+{\cal G}&=&\frac{l_1}{2\,\beta_1}\, e^{\beta_1\,u^{(1)}}\,\left(\sigma_1\,e^{\beta_1\,\lambda^{(1)}}+\frac{1}{\sigma_1}\,e^{-\beta_1\,\lambda^{(1)}}\right)\nonumber\\
&&\phantom{m}+
\frac{l_2}{2\,\beta_2}\, e^{\beta_2\,u^{(2)}}\,\left(\sigma_2\,e^{\beta_2\,\lambda^{(2)}}+\frac{1}{\sigma_2}\,e^{-\beta_2\,\lambda^{(2)}}\right)\nonumber
\\
&&\phantom{mm}+\frac{l_3}{2\,\beta_3}\, \, e^{\beta_3\,u^{(3)}}\,\left(\sigma_3\,e^{\beta_3\,(-a\,\lambda^{(1)}-c\,\lambda^{(2)})}+\frac{1}{\sigma_3}\,e^{-\beta_3\,(a\,\lambda^{(1)}+c\,\lambda^{(2)})}\right)\nonumber\\
&&\phantom{mmm}+\frac{l_4}{2\,\beta_4}\, e^{\beta_4 \,u^{(4)}}\,\left(\sigma_4\,e^{\beta_4\,(-c\,\lambda^{(1)}+a\,\lambda^{(2)})}+\frac{1}{\sigma_4}\,e^{-\beta_4\,(c\,\lambda^{(1)}-a\,\lambda^{(2)})}\right).
\end{eqnarray}
On the other hand, by setting $a_1=\beta_1,$ $a_2=2\,\beta_1,$ $b_1=2\beta_2,$ $b_2=\beta_2,$   it is found that
$$\gamma_1^T=\beta_1\,(1,2),\quad \gamma_2^T=\beta_2\,(2,1),\quad \gamma_3^T=\beta_3\,(a-2c, c-2a),\quad \gamma_4^T=\beta_4\,(c+2a,\, -a-2c),$$
$$\delta_1^T=\beta_1\,(1,-2),\quad \delta_2^T=\beta_2\,(-2,\,1),\quad \delta_3^T=\beta_3\,(a+2c,\, c+2a),\quad \delta_4^T=\beta_4\,(c-2a, -a+2c)$$
and the defect potential becomes
\begin{eqnarray}
&&{\cal D}={\cal F}+{\cal G}=\frac{l_1}{2\,\beta_1}\, e^{\beta_1u^{(1)}}\,\left(\sigma_1\,e^{\beta_1\,V^{(1)}}+
\frac{1}{\sigma_1}\,e^{-\beta_1\,V^{(1)}}\right)\nonumber\\
&&\phantom{=F+G=mm}+\frac{l_2}{2\,\beta_2}\, e^{\beta_2u^{(2)}}\,\left(\sigma_2\,e^{\beta_2\,V^{(2)}}+\frac{1}{\sigma_2}\,e^{-\beta_2\,V^{(2)}}\right)\nonumber\\
&&\phantom{=F+G=mmm}+\frac{l_3}{2\,\beta_3}\, \, e^{\beta_3u^{(3)}}\,\left(\sigma_3\,e^{\beta_3\,V^{(3)}}+\frac{1}{\sigma_3}\,e^{-\beta_3\,V^{(3)}}\right)\nonumber\\
&&\phantom{=F+G=mmmm}+\frac{l_4}{2\,\beta_4}\, e^{\beta_4u^{(4)}}\,\left(\sigma_4\,e^{\beta_4\,V^{(4)}}+\frac{1}{\sigma_4}\,e^{-\beta_4\,V^{(4)}}\right),
\end{eqnarray}
with
\begin{eqnarray*}
&&V^{(1)}= u^{(2)}+c \, u^{(3)}-a\,u^{(4)}+\lambda^{(1)}   ,\\
&&V^{(2)}=u^{(1)}+a\,u^{(2)}-c\,u^{(4)}+\lambda^{(2)}   ,\\
&&V^{(3)}=-c\,u^{(1)}-a\,u^{(2)}-2\,a\,c\,u^{(3)}+(a^2-c^2)\,u^{(4)}-a\,\lambda^{(1)}-c\,\lambda^{(2)},\\ &&V^{(4)}=\phantom{-}a\,u^{(1)}-c\,u^{(2)}+(a^2-c^2)\,u^{(3)}+2\,a\,c\,u^{(4)}-c\,\lambda^{(1)}+a\,\lambda^{(2)}.
\end{eqnarray*}
Note, by setting $c=1$ the matrix $C$ becomes diagonal and the defect potential simplifies further. It is clear that the matrix $C$ controls the way in which the Liouville fields interact with the auxiliary fields $\lambda^{(i)}$ and amongst themselves at the junction. The mixing amongst the Liouville fields is determined by the relations \eqref{LiouvilleMixing}.

\p On the other hand, an attempt with the sine-Gordon model proves to be fruitless. In fact, taking the sine-Gordon potential for the $u^{(i)}$ field to be
$$U^{(i)}=\frac{m_i^2}{2\beta_i^2 c_i}\left(e^{\beta_i\sqrt{c_i}u^{(i)}}+e^{-\beta_i\sqrt{c_i}u^{(i)}}\right),$$
and the same $C$ matrix used in the Liouville case, the right hand side of \eqref{FGquadratic} becomes
\begin{equation}\label{SGRHSQuadraticR}
2\,\sum_{j=1}^2(\epsilon_i\,c_i\,U^{(i)}
+\tilde{\epsilon_i}\,\tilde{c_i}\,\tilde{U^{(i)}})=\sum_{k=1}^4\,\omega_k\,\left(\frac{m_k}{\beta_k}\right)^2\,\left(e^{\alpha_k\cdot (p+\omega_k q)}+e^{-\alpha_k\cdot (p+\omega_k q)}\right).
\end{equation}
Suitable expressions for the functions ${\cal F}$ and ${\cal G}$ are:
$${\cal F}=\sum_{k=1}^4\,e^{\alpha_k\cdot (p+\lambda)}\,x_k(q)\,
+e^{-\alpha_k\cdot (p+\lambda)}\,\hat{x}_k(q),
\quad {\cal G}=\sum_{l=1}^4\,e^{\alpha_l\cdot (p-\lambda)}\,y_l(q) +
e^{-\alpha_l\cdot (p-\lambda)}\,\hat{y}_l(q)$$
where $x_i,$ $y_i$, $\hat{x}_i,$ $\hat{y}_i,$ are unspecified functions of $q.$ The left hand side of \eqref{FGquadratic} reads

\begin{eqnarray*}
&& \sum_{i=1}^2\left(\frac{\partial {\cal D}}{\partial q^{(i)}}\,\frac{\partial {\cal P}}{\partial \lambda^{(i)}}
-\frac{\partial {\cal P}}{\partial q^{(i)}}\,\frac{\partial {\cal  D}}{\partial \lambda^{(i)}}\right)\\
&&=
2\,\sum_{k,l=1}^4\,\left[e^{(\alpha_k+\alpha_l)\cdot p/2}\left(e^{(\alpha_k-\alpha_l)\cdot\lambda/2}\,y_l\,\left(\alpha_l\cdot\frac{\partial x_k}{\partial  q}\right)+
e^{-(\alpha_k-\alpha_l)\cdot\lambda/2}\,x_l\,\left(\alpha_l\cdot\frac{\partial y_k}{\partial q}\right)\right)\right.\\
&&\phantom{m}\left.-e^{-(\alpha_k+\alpha_l)\cdot p/2}\left(e^{-(\alpha_k-\alpha_l)\cdot\lambda/2}\,\hat{y}_l\,\left(\alpha_l\cdot\frac{\partial \hat{x}_k}{\partial q}\right)+
e^{(\alpha_k-\alpha_l)\cdot\lambda/2}\,\hat{x}_l\,\left(\alpha_l\cdot\frac{\partial \hat{y}_k}{\partial q}\right)\right)\right.\\
&&\phantom{mm}\left.-e^{(\alpha_k-\alpha_l)\cdot p/2}\left(e^{(\alpha_k+\alpha_l)\cdot\lambda/2}\,\hat{y}_l\,\left(\alpha_l\cdot\frac{\partial x_k}{\partial q}\right)+
e^{-(\alpha_k+\alpha_l)\cdot\lambda/2}\,\hat{x}_l\,\left(\alpha_l\cdot\frac{\partial y_k}{\partial q}\right)\right)\right.\\
&&\phantom{mmm}\left.+e^{-(\alpha_k-\alpha_l)\cdot p/2}\left(e^{-(\alpha_k+\alpha_l)\cdot\lambda/2}\,y_l\,\left(\alpha_l\cdot\frac{\partial \hat{x}_k}{\partial q}\right)+
e^{(\alpha_k+\alpha_l)\cdot\lambda/2}\,x_l\,\left(\alpha_l\cdot\frac{\partial \hat{y}_k}{\partial q}\right)\right)\right].
\end{eqnarray*}
Terms for which $k=l$ must be equal to \eqref{SGRHSQuadraticR}. Hence, the following constraints follow

\begin{eqnarray}
\omega_k\,\left(\frac{m_k}{\beta_k}\right)^2\,e^{\omega_k\,\alpha_k\cdot q}&\equiv&
2\left(y_k\,\left(\alpha_k\cdot \frac{\partial x_k}{\partial q}\right)+
x_k\,\left(\alpha_k\cdot\frac{\partial y_k}{\partial q}\right)\right),\label{SGconstraint_1}\\
\omega_k\,\left(\frac{m_k}{\beta_k}\right)^2\,e^{-\omega_k\,\alpha_k\cdot q}&\equiv&
-2\left(\hat{y}_k\,\left(\alpha_k\cdot\frac{\partial\hat{x}_k}{\partial q}\right)+
\hat{x}_k\,\left(\alpha_k\cdot \frac{\partial \hat{y}_k}{\partial q}\right)\right),\label{SGconstraint_2}\\
\hat{y}_k\,\left(\alpha_k\cdot \frac{\partial x_k}{\partial q}\right)&\equiv&
x_k\,\left(\alpha_k\cdot\frac{\partial \hat{y}_k}{\partial q}\right),\label{SGconstraint_3}\\
\hat{x}_k\,\left(\alpha_k\cdot \frac{\partial y_k}{\partial q}\right)&\equiv&
y_k\,\left(\alpha_k\cdot\frac{\partial \hat{x}_k}{\partial q}\right).\label{SGconstraint_4}
\end{eqnarray}
A few manipulations  on these equations, reveal that they do not have a solution. In fact, using \eqref{SGconstraint_3} and \eqref{SGconstraint_4} into \eqref{SGconstraint_2}, the latter becomes
$$\omega_k\,\left(\frac{m_k}{\beta_k}\right)^2\,e^{-\omega_k\,\alpha_k\cdot q}\,(x_k\,y_k)(\hat{x}_k\,\hat{y}_k)^{-1}=
-2\left(y_k\,\left(\alpha_k\cdot \frac{\partial x_k}{\partial q}\right)+
x_k\,\left(\alpha_k\cdot\frac{\partial y_k}{\partial q}\right)\right).$$
Combining this expression with \eqref{SGconstraint_1}, the following  constraint is obtained
$$e^{-2\omega_k\,\alpha_k\cdot q}\,(x_k\,y_k)(\hat{x}_k\,\hat{y}_k)^{-1}= -1,\quad \rightarrow\quad \hat{x}_k\,\hat{y}_k=-(x_k\,y_k)\,e^{-2\omega_k\,\alpha_k\cdot q}.$$
This expression can be used inside \eqref{SGconstraint_2}, since \eqref{SGconstraint_2} can be rewritten as follows
$$\omega_k\,\left(\frac{m_k}{\beta_k}\right)^2\,e^{-\omega_k\,\alpha_k\cdot q}=
-2\,\left(\alpha_k\cdot \frac{\partial( \hat{x}_k\,\hat{y}_k)}{\partial q}\right).$$
By expanding it, it is found that
$$2\,\left(\alpha_k\cdot \frac{\partial( x_k\,y_k)}{\partial q}\right)=\omega_k\,\left(\frac{m_k}{\beta_k}\right)^2\,e^{\omega_k\,\alpha_k\cdot q}+2\,\omega_k\,( x_k\,y_k)\,(\alpha_k\cdot\alpha_k).$$
This expression can be compared with \eqref{SGconstraint_1}, which rewritten reads
$$2\,\left(\alpha_k\cdot \frac{\partial( x_k\,y_k)}{\partial q}\right)=\omega_k\,\left(\frac{m_k}{\beta_k}\right)^2\,e^{\omega_k\,\alpha_k\cdot q},$$
implying $( x_k\,y_k)=0,$ which clearly is not a possibility.

\p The only way out is the possibility of having two pairs of type II defects. This possibility arises by noticing that the matrix $C$ simplifies by setting $a=0,$ $c=-1.$ The advantage of this situation is that by setting $\beta_1=\beta_4$ and $\beta_2=\beta_3,$ it can be noticed that $\alpha_1=\alpha_4$ and $\alpha_2=\alpha_3.$ Then the constraints analogous to \eqref{SGconstraint_1}-\eqref{SGconstraint_4} become:
\begin{eqnarray*}
\left(\frac{m_k}{\beta_k}\right)^2\,\left(e^{\alpha_k\cdot q}-e^{-\alpha_k\cdot q}\right)&\equiv&
2\left(y_k\,\left(\alpha_k\cdot\frac{\partial x_k}{\partial q}\right)+
x_k\,\left(\alpha_k\cdot\frac{\partial y_k}{\partial q}\right)\right),\\
\left(\frac{m_k}{\beta_k}\right)^2\,\left(e^{\alpha_k\cdot q}-e^{-\alpha_k\cdot q}\right)&\equiv&
2\left(\hat{y}_k\,\left(\alpha_k\cdot \frac{\partial \hat{x}_k}{\partial q}\right)+
\hat{x}_k\,\left(\alpha_k\cdot\frac{\partial \hat{y}_k}{\partial q}\right)\right),\\
\hat{y}_k\,\left(\alpha_k\cdot\frac{\partial x_k}{\partial q}\right)&\equiv&
x_k\,\left(\alpha_k\cdot\frac{\partial \hat{y}_k}{\partial q}\right),\\
\hat{x}_k\,\left(\alpha_k\cdot\frac{\partial y_k}{\partial q}\right)&\equiv&
y_k\,\left(\alpha_k\cdot\frac{\partial \hat{x}_k}{\partial q}\right),\
\end{eqnarray*}
where $k=1,2,$ only. The solution of these constraints leads to the familiar result.

\section{Concluding remarks}

\p In this article, the existence of integrable defects able to relate domains with different wave speeds has been explored. For the sine-Gordon model, a junction can only have two branches,  or be a junction that acts as a meeting point of defects, each with two branches. In this case the junction allows to pair branches in several ways and effectively acts as a switch and/or a store of topological charge. Moreover, though the coupling constants of the models in the two domains of an $N=2$ junction are different, the $S$ matrices appropriate to each branch are the same since the ratio $c/\beta^2$ is preserved across the junction. On the other hand, if the fields on the different branches are free massive fields, Liouville or free massless, the integrable defects can support multiple branches at the junction provided there is an even number of them.  In this case, the fields interact at the junction and their mixing is controlled by a variety of parameters. Multi-field generalisations (for example the conformal or affine Toda field theories) have not been considered in this first analysis but should be the subject of future investigation.

\p An alternative setup has been proposed by Sobirov et al. in ref\cite{sobirov2015}. There, the fields on each branch of a network or graph are required to be continuous at the junction so that, evaluated at the junction, $u^{(1)}=u^{(2)}=\dots =u^{(N)}$. Then, in the notation of the present article, \eqref{Edot}, the conservation of energy requires
$$\dot\ce=\sum_{i=1}^N \epsilon_ic_i^2\left[u^{(i)}_tu^{(i)}_x\right]_{x_0}=\left[u^{(1)}_t\sum_{i=1}^N\epsilon_ic_i^2u^{(i)}_x\right]_{x_0}=0,$$
and thus at the junction
\begin{equation}\label{scondition1}
\sum_{i=1}^N\epsilon_ic_i^2u^{(i)}_x=0.
\end{equation}
On the other hand, conserving momentum (term by term in the expression for $\dot\cm$ in this setup) is guaranteed by setting
\begin{equation}\label{scondition2}
\sum_{i=1}^N\epsilon_ic_i=0,\quad \sum_{i=1}^N\epsilon_ic_i^3(u^{(i)}_x)^2=0, \quad U^{(1)}=U^{(2)}=\dots=U^{(N)}.
\end{equation}
The two conditions involving the spatial derivatives of the fields on each branch evaluated at the junction are then solved together by taking
\begin{equation}\label{scondition3}
c_1u^{(1)}_x=c_2u^{(2)}_x=\dots=c_Nu^{(N)}_x.
\end{equation}
Note also, since the potentials must be the same along each branch,  the coupling constants must be identical. The conditions \eqref{scondition1}, \eqref{scondition2}, \eqref{scondition3} are applicable to any choice of $N$. However, if $N=2$, the junction disappears because the wave speeds on either side of it must be the same. Note also, this type of junction introduces no additional tunable parameters at the junction.

\p It is worth noting that if it became necessary to assemble junctions joining branches with different wave speeds then it would always be possible to change speed along a particular branch by inserting a type I defect, in the manner explained in section 2, at the junction of the two domains along the branch. For example, taking $N=3$, $\epsilon_1=-1,\ \epsilon_2=\epsilon_3=1$, inserting defects in the branches 2 and 3 to restore the wave speeds along those branches to $c_1$, and using \eqref{kappaequation}, requires
$$\frac{c_1}{\beta_2^2}=\frac{c_2}{\beta_1^2},\ \ \frac{c_1}{\beta_3^2}=\frac{c_3}{\beta_1^2}.$$
Hence, the new coupling constants along the branches 2 and 3 should be chosen so that
$$\frac{1}{\beta_1^2}=\frac{1}{\beta_2^2}+\frac{1}{\beta_3^2},$$
the latter following from the first constraint in \eqref{scondition2}.

\p Though this article has focussed on the classical properties of field theories defined on branches meeting at junctions, it would be interesting also to explore if and how a network might support quantum field theories that interact consistently at its junctions. The purpose would be to find generalisations of  ideas contained, for example, in \cite{kostrykin1999, Bellazzini2006}. In this context it is not precisely clear what the consistency condition might be. In the type of junction discussed in \cite{sobirov2015}, the sine-Gordon coupling constant $\beta$ on each branch is the same but the wave speeds in the three branches add appropriately. This means that the dimensionless quantities
$$\frac{c}{\hbar \beta^2}$$
 are different on each of the branches and therefore the S-matrices have the same form but with different parameters. This contrasts with the situation discussed earlier  in  section \ref{Remarks}. In the notation of \cite{bcz2005}, the Zamolodchikov sine-Gordon S-matrix \cite{Zam79} depends on the coupling constant via
$$q=e^{i\pi\gamma},\ \ \gamma=\frac{8\pi c}{\hbar\beta^2}-1,\ \ x=e^{\gamma\theta}.$$
Thus, in a situation at a junction where the wave speeds on the three branches satisfy $c_1=c_2+c_3$, the following relations hold for a specific rapidity:
$$q_2q_3=-q_1,\ \ x_2x_3=x_1\,e^{-\theta}.$$
If two solitons with differing rapidities approach the junction along branch $1$ then they may scatter before or after splitting at the junction. Hence, adapting the Yang-Baxter equation to this situation suggests a suitable consistency condition might then be
\begin{equation}\label{junctionYB}S_{1\,ab}^{\phantom{0\,}dc}(\theta_{12}) J_d^{ef}(\theta_1)J_c^{gh}(\theta_2)=J_b^{pu}(\theta_2)J_a^{qv}(\theta_1)S_{2\,qp}^{\phantom{1\,}ge}(\theta_{12}) S_{3\,vu}^{\phantom{2\,}hf}(\theta_{12}), \ \ \theta_{12}=\theta_1-\theta_2,
\end{equation}
where repeated indices are summed. Here, the rapidity-dependant  quantities with three labels, $J_a^{bc}(\theta)$ represent the splitting at the junction. It is straightforward to see that if the junction strictly preserves topological charge as claimed for instance in \cite{sobirov2015}, then there is no solution to \eqref{junctionYB}. However, it is less clear if there are solutions to a modified version of \eqref{junctionYB} if topological charge is not preserved and may be deposited on or removed from a junction, as  happens for integrable defects. Investigating equations of type \eqref{junctionYB} will be the focus of  a future discussion.

\section{Acknowledgements}

We are grateful to Robert Parini for  discussions on this topic and for bringing the paper \cite{sobirov2015} to our attention.

\vfill\newpage

\noindent\textbf{\large Appendices}
  \vspace{-20pt}
  \appendix
  \section{Type II junction - the setting}

Consider a junction with $N$ branches located at $x=0$. The starting point for a type II junction could be the following junction Lagrangian:
\begin{eqnarray*}
{\cal L}&=&\left(\sum_{i,j=1}^N\, u^{(i)}A_{ij}u_t^{(j)}+\sum_{k,l=1}^M\, \lambda^{(k)}G_{kl}\lambda_t^{(l)}+\sum_{i=1}^N\, \sum_{k=1}^M\, u^{(i)}C_{ik}\lambda_t^{(k)}-{\cal D}\left(u^{(i)},\lambda^{(k)}\right)\right)\delta(x),\\
&&\phantom{mmmmmmm}A^T=-A,\quad G^T=-G,
\end{eqnarray*}
where $M$ is the number of the auxiliary fields $\lambda^{(k)},$ which represent the additional degrees of freedom at the junction and ${\cal D}$ is the potential at the junction, which depends on all fields in the systems, i.e. $u^{(i)}$ and $\lambda^{(k)}.$ Since they are free, it is possible to perform transformations which allow to simplify the original setting. By following the manipulations in \cite{br2017}, it can be shown that the auxiliary fields $\lambda^{(k)}$ can be split into two different sets, $\mu^{(k)}$ and $\lambda^{(k)}.$ The interesting feature is that fields  $\mu^{(k)}$ only interact with themselves, at the junction, while fields $\lambda^{(k)}$ interact exclusively with the fields $u^{(i)}.$ In summary, the new junction Lagrangian becomes
\begin{eqnarray*}
{\cal L}&=&\left(\sum_{i,j=1}^N\,\frac{1}{2}\,u^{(i)}A_{ij}u_t^{(j)}+\sum_{k,l=1}^{M_1}\,\frac{1}{2}\,\mu^{(k)}G_{kl}\mu_t^{(l)}+\sum_{i=1}^N\, \sum_{m=1}^{M_2}\,u^{(i)}C_{im}\lambda_t^{(m)}\right.\\
&&\left.\phantom{mmm\sum_{i,j=1}^N\,\frac{1}{2}}-{\cal D}(u^{(i)},\mu^{(k)},\lambda^{(m)})\right)\delta(x),
\end{eqnarray*}
with $M_1+M_2=M$ and, in addition, $G^TG=I,$ $G^2=-I.$ Note that the presence of the fields $\mu^{(k)}$ implies the presence of an $A$ matrix.
\p The conditions at the junction are:
\begin{eqnarray}
c_i^2u_x^{(i)}&=&\left(\sum_{j=1}^N\,A_{ij}u_t^{(j)}+\sum_{m=1}^{M_2}\,C_{im} \lambda_t^{(m)}-\frac{\partial {\cal D}}{\partial u^{(i)}}\right)\,\epsilon_i,\quad i=1,\dots,N,\label{JunctionConditionsAppendix1}\\
 \sum_{l=1}^{M_1}\,G_{kl}\,\mu_t^{(l)}&=&\frac{\partial {\cal D}}{\partial \mu^{(k)}},\quad i=1,\dots, M_1,\label{JunctionConditionsAppendix2}\\
\sum_{i=1}^{N}\,u_t^{(i)}\,C_{im}&=&-\frac{\partial {\cal D}}{\partial \lambda^{(m)}},\quad m=1,\dots, M_2.\label{JunctionConditionsAppendix3}
\end{eqnarray}
Then, concerning the energy ${\cal E}$
\begin{equation*}
\dot {\cal E}=\sum_{i=1}^N\,\epsilon_i\,c_i^2\,u_t^{(i)}u_x^{(i)}=-\sum_{i=1}^N\,\left(u_t^{(i)}\,\frac{\partial {\cal D}}{\partial u^{(i)}}\right)-\sum_{m=1}^{M_2}\,\left(\lambda_t^{(i)}\,\frac{\partial {\cal D}}{\partial \lambda^{(i)}}\right)=\frac{d {\cal D}}{d t}.
\end{equation*}
On the other hand, applying condition \eqref{JunctionConditionsAppendix1} on the time derivative of the scaled momentum, i.e.
\begin{eqnarray*}
\dot {\cal M}&=&\sum_{i=1}^N\,\left(\frac{1}{2}(u_t^{(i)})^2+\frac{c_i^2}{2}(u_x^{(i)})^2-U^{(i)}\right)\,\epsilon_i\,c_i,
\end{eqnarray*}
a constraint can be derived staright away by looking at the coefficients of the terms square in the time derivative of the auxiliary fields. They need to cancel, which implies
\begin{equation}\label{constraintAppendix1}
C^T\,\epsilon\,c^{-1}C=0.
\end{equation}
Note that, $C$ is an $N\times M_2$ matrix, where $M_2\leq N$ is the number of $\lambda^{(m)}$ fields.
A suitable transformation on the fields $\lambda^{(m)}$ can be used to reduce
the number of linearly independent columns of $C$ so that all the other columns are zero. This means that there will be only $n\leq M_2$ $\lambda^{(m)}$ since the other $M_2-n$ decouple and can be removed altogether from the junction Lagrangian.
Let us call $n$ the number of linearly independent columns of $C$. Since the matrix $C$ has rank $n,$ it follows that it must exist a square $(n\times n)$ sub-matrix, which is invertible.
Hence $C$ can be split into an $(n\times n)$ invertible matrix and
an $((N-n) \times  n)$ matrix. Note that, in addition,  a suitable transformation on the field $\mu^{(m)}$
can be performed to reduce the $(n\times n)$ invertible matrix to the $(n\times n)$ identity.
It is worth pointing out that the limiting case $n=N,$ allows to set $C=1.$ As a consequence,
the constraint \eqref{constraintAppendix1} reduces to
$$\epsilon\,c^{-1}=0$$
and it cannot clearly be satisfied.  Note that for $N=2$ and $n=1,$ i.e. one auxiliary field, the matrix $C$ is:
$$C=\left(
     \begin{array}{c}
       1 \\
       C_{21} \\
     \end{array}
   \right)
$$
and the constraint \eqref{constraintAppendix1} becomes
$$\left(C^T \epsilon\,c^{-1} C\right)_{11}=0 \quad \rightarrow \quad \epsilon_1=-\epsilon_2,\quad C_{21}=\pm \sqrt{\frac{c_2}{c_1}}.$$
Hence the familiar type II defect is recovered.

\subsection{The constraints}

\p In view of what has been found, it is convenient to split the fields $u^{(i)}$ into two separated groups, according to the way in which they interact with the $\mu^{(k)}$ fields, that is via the diagonal part of the $C$ matrix or the remaining one. Hence, the new starting point for the junction part of the Lagrangian is:

\begin{eqnarray*}
{\cal L}&=&\left(
\sum_{i,j=1}^n\,u^{(i)}\frac{A_{ij}}{2}u_t^{(j)}+\sum_{k,l=1}^{N-n}\,\tilde{u}^{(k)}
\frac{\tilde{A}_{kl}}{2}\tilde{u}_t^{(l)}+\sum_{i=1}^n\,\sum_{k=1}^{N-n}\,
u^{(i)}B_{ik}\tilde{u}_t^{(k)}+\sum_{i,j=1}^n\,u^{(i)}\delta_{ij}\lambda_t^{(j)}\right.\\
&&\phantom{mmmm}\left.+\sum_{j=1}^n\,\sum_{k=1}^{N-n}\,\tilde{u}^{(k)}\tilde{C}_{kj}\lambda_t^{(j)}- {\cal D}\left(u^{(i)}, \tilde{u}^{(k)}, \lambda^{(j)}\right)\right)\,\delta(x),
\end{eqnarray*}
where, for simplicity, the fields $\mu^{(m)}$ have been removed altogether.
Note that $A=-A^T$ is a  $(n\times n)$ matrix, $\tilde{A}=-\tilde{A}^T$ a $((N-n)\times (N-n))$  matrix,
$B$ is a $(n\times (N-n))$ and $\tilde{C}$ is a $((N-n)\times n)$ matrix.
The conditions at the junction are:
\begin{eqnarray}
&&c_i^2u_x^{(i)}=\left(\sum_{j=1}^n\,\left(A_{ij}u_t^{(j)}+\delta_{ij} \lambda_t^{(j)}\right)+ \sum_{k=1}^{N-n}\,B_{ik}\tilde{u}_t^{(k)}-\frac{\partial {\cal D}}{\partial u^{(i)}}\right)\,\epsilon_i,\  i=1,\dots,n,\label{JunctionNewConditionsAppendix1}\\
&&\tilde{c}_k^2\tilde{u}_x^{(k)}=\left(\sum_{k=1}^{N-n}\,\tilde{A}_{kl}\tilde{u}_t^{(l)}+\sum_{j=1}^n\,\left(\tilde{C}_{kj} \lambda_t^{(j)}- B_{jk}u_t^{(j)}\right)-\frac{\partial {\cal D}}{\partial \tilde{u}^{(k)}}\right)\,\tilde{\epsilon}_k,\  k=1,\dots,N-n,\ \ \  \ \label{JunctionNewConditionsAppendix2}\\
 &&\phantom{mmmmmm}u_t^{(i)}\,\delta_{ji}+\sum_{k=1}^{N-n}\,\tilde{u}_t^{(k)}\,\tilde{C}_{ki}=-\frac{\partial {\cal D}}{\partial \lambda^{(i)}},\  i=1,\dots,n.\label{JunctionNewConditionsAppendix3}
\end{eqnarray}
It is worth looking afresh at the time derivative of the scaled momentum. All the constraints derived below, are due to the fact that the coefficients of certain terms containing combinations of different fields must cancel independently. First, by implementing the relations \eqref{JunctionNewConditionsAppendix1} and \eqref{JunctionNewConditionsAppendix2}, a constraint is immediately derived from the coefficients quadratic in the auxiliary fields $ \lambda_t^{(j)}.$ They lead to:
\begin{equation}\label{constraintNewAppendix1}
{\epsilon}{c^{-1}}+\tilde{C}^T\,{\tilde{\epsilon}}{\tilde{c}}^{-1}\,\tilde{C}=0.
\end{equation}
Second, using the condition \eqref{JunctionNewConditionsAppendix3} together with the constraint already obtained, it is possible to get two further constraints. The former is derived from the coefficients of the terms proportional to the product of $\tilde{u}_t^{(k)}$ and $\lambda_t^{(j)}$ fields. The constraint derived is:
\begin{equation}\label{constraintNewAppendix2}
\tilde{A}+\tilde{C}A\tilde{C}^T=\tilde{C}B-(\tilde{C}B)^T.
\end{equation}
The latter stems from the terms quadratic in the fields $\tilde{u}_t^{(k)}.$ This
constraint is:
\begin{equation}\label{constraintNewappendix3}
\tilde{\epsilon}\,\tilde{c}+\tilde{C}\,\epsilon\,c\,\tilde{C}^T=0.
\end{equation}

\p The remaining terms lead to
$$\dot {\cal M}={\cal X}+{\cal Y}=-\frac{d{\cal P}}{dt}\,,$$
with
\begin{eqnarray*}
&&{\cal X}=\sum_{i=1}^n\frac{\epsilon_i}{2\,c_i}\left(\frac{\partial {\cal D}}{\partial \,u^{(i)}}\right)^2+\sum_{k=1}^{N-n}\frac{\tilde{\epsilon}_k}{2\,\tilde{c}_k}\left(\frac{\partial {\cal D}}{\partial \,\tilde{u}^{(k)}}\right)^2\phantom{mmm}\\
&&\phantom{mm}+\sum_{i,j=1}^n\frac{1}{2}\,\frac{\partial {\cal D}}{\partial \,\lambda^{(i)}}\left(\epsilon\,c-A\,{\epsilon}{c^{-1}}\,A+
B\,{\tilde{\epsilon}}{\tilde{c}}^{-1}\,B^T\right)_{ij}\,\frac{\partial {\cal D}}{\partial \,\lambda^{(j)}}\\
&&\phantom{mmm}-\sum_{i,j=1}^n\frac{\partial {\cal D}}{\partial \,\lambda^{(i)}}\,\left(A\,{\epsilon}{c^{-1}}\right)_{ij}\frac{\partial {\cal D}}{\partial \,u^{(j)}}
-\sum_{j=1}^n\sum_{k=1}^{N-n}\frac{\partial {\cal D}}{\partial \,\lambda^{(j)}}\,\left(B\,{\tilde{\epsilon}}{\tilde{c}}^{-1}\right)_{jk}\frac{\partial {\cal D}}{\partial \,\tilde{u}^{(k)}}\\
&&\phantom{mmmmm}-\sum_{i=1}^n\epsilon_i\,c_iU^{(i)}
-\sum_{k=1}^{N-n}\tilde{\epsilon}_k\,\tilde{c}_k\tilde{U}^{(k)},
\end{eqnarray*}
and
\begin{eqnarray*}
&&{\cal Y}=-\sum_{j=1}^n\,\sum_{k=1}^{N-n}\,\lambda_t^{(j)}\left(\tilde{C}^T\,\tilde{\epsilon}\tilde{c}^{-1}\right)_{jk}\frac{\partial {\cal D}}{\partial \tilde{u}^{(k)}}-\sum_{i,j=1}^n\,\lambda_t^{(i)}\left(\epsilon c^{-1}\right)_{ij}\,\frac{\partial {\cal D}}{\partial u^{(j)}}\\
&&\phantom{mmmm}-\sum_{j=1}^n\,\sum_{k=1}^{N-n}\,\tilde{u}^{(k)}\,\left(B^T\,\epsilon c^{-1} +\tilde{C}\,A\,\epsilon c^{-1}\right)_{kj}\,\frac{\partial {\cal D}}{\partial \,u^{(j)}}
\nonumber \\ &&\phantom{mmmm}+ \sum_{k,l=1}^{N-n}\,\tilde{u}^{(k)}\,\left(\tilde{A}\,\tilde{\epsilon}\tilde{c}^{-1}-\tilde{C}\,B\,\tilde{\epsilon}\tilde{c}^{-1}\right)_{kl}\,\frac{\partial {\cal D}}{\partial \,\tilde{u}^{(l)}}\\
&&\phantom{mmmm}+\sum_{j=1}^n\,\sum_{k=1}^{N-n}\,\tilde{u}_t^{(k)}\left(\tilde{C}\,\epsilon c -\tilde{C}\,A\,\epsilon c^{-1}\,A+\tilde{C}\,B\,\tilde{\epsilon}\tilde{c}^{-1}\,B
\right. \nonumber \\ &&\left. \phantom{mmmmmmmmmmmmmm} -B^T\,\epsilon c^{-1}\,A -\tilde{A}\,\tilde{\epsilon}\tilde{c}^{-1}\,B^T\right)_{kj}\,
\frac{\partial {\cal D}}{\partial \lambda^{(j)}}.
\end{eqnarray*}

\section{Type II junction - free fields}

\p Consider the matrix $C$ given by \eqref{C_solutionDiagonal}.  Several solutions can be found, for instance
\begin{eqnarray}
F_1&=&\left(\begin{array}{cc}
a_1 & \alpha \\
\alpha & a_2
\end{array}\right),\quad F_2=\frac{m^2}{b_1b_2-\beta^2}\left(\begin{array}{cc}
b_2 & -\beta \\
 -\beta & b_1
\end{array}\right),\quad F_3=0,\\
G_1&=&\left(\begin{array}{cc}
b_1 & \beta \\
\beta & b_2
\end{array}\right),\quad G_2=\frac{m^2}{a_1a_2-\alpha^2}\left(\begin{array}{cc}
a_2 & -\alpha \\
 -\alpha & a_1
\end{array}\right),\quad G_3=0,
\end{eqnarray}
with $m=m_1=m_2,$
or
\begin{eqnarray}
F_1&=&\left(\begin{array}{cc}
a_1 & 0 \\
0 & a_2
\end{array}\right),\quad F_2=\left(\begin{array}{cc}
f_1 & 0 \\
0 & f_2
\end{array}\right),\quad F_3=\left(\begin{array}{cc}
0 & \sigma_1 \\
\sigma_2 & 0
\end{array}\right),\\
G_1&=&\left(\begin{array}{cc}
b_1 &0 \\
0 & b_2
\end{array}\right),\quad G_2=\left(\begin{array}{cc}
g_1 & 0 \\
0 & g_2
\end{array}\right),\quad G_3=\left(\begin{array}{cc}
0 & \omega_1 \\
\omega_2 & 0
\end{array}\right),
\end{eqnarray}
with
\begin{eqnarray*}
\frac{\sigma_2}{\sigma_1}&=&\frac{a_2\,g_1}{b_1\,f_1}=\frac{b_2\,f_2}{a_1\,g_2},\quad   g_1\,a_1+\omega_2\,\sigma_1=b_1\,f_1+\omega_1\,\sigma_2=m_1^2,\\
\frac{\omega_2}{\omega_1}&=&\frac{a_2\,g_2}{b_1\,f_2}=\frac{b_2\,f_1}{a_1\,g_1},
\quad  g_2\,a_2+\omega_1\,\sigma_2=b_2\,f_2+\omega_2\,\sigma_1=m_2^2.
\end{eqnarray*}
In the most general case, \eqref{C_solution}, the first solution still holds. On the other hand, the second solution holds only if the two masses are set to be equal. Hence, the constraints become
\begin{eqnarray*}
&&\frac{\omega_1}{\omega_2}=\frac{f_2}{f_1},\quad \frac{\sigma_1}{\sigma_2}=\frac{g_2}{g_1},\quad \frac{b_2}{a_1}=\frac{g_1}{f_2},\quad \frac{b_1}{a_2}=\frac{g_2}{f_1},\\
&&\quad  g_2\,a_2+\omega_1\,\sigma_2= g_1\,a_1+\omega_2\,\sigma_1=m^2,
\end{eqnarray*}
where $m=m_1=m_2.$ In order to calculate the transmission matrix, $T,$ corresponding to plane waves on the four legs, the sewing conditions at the junction, \eqref{JunctionNewConditions1}-\eqref{JunctionNewConditions3}, must be used. When rewritten in terms of variables $p$ and $q$  defined  in \eqref{New_q_and_p}, they are
\begin{eqnarray*}
&&2\,\lambda_t^{(i)}-\sum_{j=1}^2\,(\epsilon\,c+{\cal A})_{ij}\,q_x^{(j)}-\sum_{j=1}^2\,(\epsilon\,c-{\cal A})_{ij}\,\epsilon_j\,p_x^{(j)}=\frac{\partial {\cal D}}{\partial q^{(i)}},\quad i=1,2,\\
&&\phantom{mmmn}\sum_{j=1}^2\,(\epsilon\,c-{\cal A})_{ij}\,q_x^{(j)}+\sum_{j=1}^2\,(\epsilon\,c+{\cal A})_{ij}\,\epsilon_j\,p_x^{(j)}=-\epsilon_i\,\frac{\partial {\cal D}}{\partial p^{(i)}},\quad i=1,2,\\
&&\phantom{mmmm}\quad 2\,q_t^{(i)}=-\frac{\partial {\cal D}}{\partial \lambda^{(i)}},\quad i=1,2,\ \  {\rm and} \  \  {\cal A}=\left(C^{T}\,{\tilde{\epsilon}}{\tilde{c}}^{-1}\,C\right)^{-1},
\end{eqnarray*}
where the subscripts $x$ and $t$ stand for derivatives with respect to $x$ and $t,$ respectively.
Note the first two relations have been obtained by taking the sum and the difference of the relations \eqref{JunctionNewConditions1}, \eqref{JunctionNewConditions2}.
As an example, consider the first solution of this section, that is the solution with $F_3=G_3=0$, and the case in which the matrix $C$ is diagonal. In this case ${\cal A}$ reduces to
${\cal A}=-\epsilon\,\tilde{c}$ and the previous expressions become
\begin{eqnarray}\label{planeWavesTypeII}
2\,\lambda_t^{(i)}-\epsilon_i\,(c_i-\tilde{c}_i)\,q_x^{(i)}-(c_i+\tilde{c}_i)\,p_x^{(i)}&=&\sum_{j=1}^2\,(F_2+G_2)_{ij}\,q^{(j)},\nonumber\\
\epsilon_i\,(c_i+\tilde{c}_i)\,q_x^{(i)}+(c_i-\tilde{c}_i)\,p_x^{(i)}&=&-\epsilon_i\,\sum_{j=1}^2\,
\left((F_1+G_1)_{ij}\,p^{(j)}+(F_1-G_1)_{ij}\,\lambda^{(j)}\right),\phantom{mm}\nonumber\\
2\,q_t^{(i)}&=&-\sum_{j=1}^2\,\left((F_1-G_1)_{ij}\,p^{(j)}+(F_1+G_1)_{ij}\,\lambda^{(j)}\right),
\end{eqnarray}
where the label $i$ runs from $1$ to $2.$
Assume no reflection and use the following notation for the plane waves - which is similar to the notation used in section [\ref{Free fields at a type I junction}],
\begin{equation*}
u^{(i)}=u_0^{(i)}e^{i(k_i x-\omega t)},\quad \tilde{u}^{(i)}=\tilde{u}_0^{(i)}e^{i(\tilde{k}_i x-\omega t)}, \quad \omega^2=\kappa^2+m^2,
\quad  (c_i\,k_i)^2\equiv (\tilde{c}_i\tilde{k}_i)^2\equiv \kappa^2,\quad i=1,2.
\end{equation*}
For simplicity, the junction is located at $x_0=0.$ After some algebra, relations \eqref{planeWavesTypeII} provide an expression for the auxiliary fields $\lambda$ and two equivalent expressions involving only the fields $u$ and $\tilde{u}.$ The latter are:
\begin{eqnarray}
&&\sum_{j=1}^2 \left({\cal X}_{ij}\,\sqrt{c_j}\,u_0^{(j)}+{\cal Y}_{ij}\,(C_{jj}\,\sqrt{\tilde{c_j}}\,\tilde{u}_0^{(j)})\right)=0,\nonumber \\
&& \sum_{j=1}^2 \left({\cal W}_{ij}\,\sqrt{c_j}\,u_0^{(j)}+{\cal Z}_{ij}\,(C_{jj}\,\sqrt{\tilde{c_j}}\,\tilde{u}_0^{(j)})\right)=0,
\end{eqnarray}
with
\begin{eqnarray*}
{\cal X}&=&i\,(A^{-1}\,\Omega+B^{-1}\,K)-\frac{1}{2}\left(A^{-1}\,\epsilon\,B-B^{-1}\,\epsilon\,A\right),\\
\quad {\cal Y}&=&
i\,(A^{-1}\,\Omega+B^{-1}\,K)+\frac{1}{2}\left(A^{-1}\,\epsilon\,B-B^{-1}\,\epsilon\,A\right),\\
{\cal W}&=&
i\,(\epsilon \,K-A^{-1}\,\Omega\,\epsilon\,B)-\left(2\,A^{-1}\,\Omega^2-\frac{E}{2}\right),
\\
{\cal Z}&=&-
i\,(\epsilon \,K-A^{-1}\,\Omega\,\epsilon\,B)-\left(2\,A^{-1}\,\Omega^2-\frac{E}{2}\right),
\end{eqnarray*}
where
$$A=F_1+G_1,\quad B=F_1-G_1,\quad E=F_2+G_2,\quad \Omega =\mbox{diag}(\omega,\omega)
\quad K =\mbox{diag}(\kappa,\kappa).$$
It follows that transmission matrix, $T,$ is:
$$\sqrt{c_i}\,u_0^{(i)}=\sum_{j=1}^2 T_{ij}\,\sqrt{\tilde{c}_j}\,\tilde{u}_0^{(j)}, \quad T\equiv -{\cal X}^{-1}\,{\cal Y}\,C=-{\cal W}^{-1}\,{\cal Z}\,C,$$
where $\tilde{u}^{(1)},$ $\tilde{u}^{(2)}$ are taken to be the incoming waves.
Without lose of generality and looking at \eqref{C_solutionDiagonal}, it  is possible to choose $C=\epsilon=\mbox{diag}(1,1),$ then $T$ is:
\begin{equation}
T=\frac{1}{f-i\,g}\,\left(\begin{array}{cc}
a-i\,b &-i\,c\\
-i\,c & a+i\,b
\end{array}\right),\quad T\,T^\dagger=I,
\end{equation}
with
\begin{eqnarray*}
a&=&-(a_1a_2-\alpha^2)(\omega+\kappa)^2-(b_1b_2-\beta^2)(\omega-\kappa)^2+
(a_1b_2+a_2b_1-2\alpha\beta)(\omega^2-\kappa^2)\\
&&-4(a_1a_2-\alpha^2)(b_1b_2-\beta^2),\\
b&=&2(b_1-b_2)(a_1a_2-\alpha^2)\,(\omega+\kappa)-2(a_1-a_2)(b_1b_2-\beta^2)\,(\omega-\kappa),\\
c&=&
4\,\beta\,(a_1a_2-\alpha^2)\,(\omega+\kappa)-4\,\alpha\,(b_1b_2-\beta^2)\,(\omega-\kappa),\\
f&=&(a_1a_2-\alpha^2)(\omega+\kappa)^2-(b_1b_2-\beta^2)(\omega-\kappa)^2+
(a_1b_2+a_2b_1-2\alpha\beta)(\omega^2-\kappa^2)\\
&&-4(a_1a_2-\alpha^2)(b_1b_2-\beta^2),\\
g&=&2(b_1+b_2)(a_1a_2-\alpha^2)\,(\omega+\kappa)-2(a_1+a_2)(b_1b_2-\beta^2)\,(\omega-\kappa),\\
\end{eqnarray*}
Some simplified cases can be analysed in more details. For instance, setting $\alpha=\beta=0,$ the transmission matrix becomes diagonal
\begin{equation}
T=-\left(\begin{array}{cc}
t_1\, &0\\
0 & t_2\,
\end{array}\right) ,
\end{equation}
with
$$t_1=\frac{i\,\kappa(a_1+b_1 )+i\,\omega\,(a_1-b_1)-2a_1b_1}{i\,\kappa\,(a_1+b_1 )+i\,\omega\,(a_1-b_1)+2a_1b_1},\quad t_2=\frac{i\,\kappa\,(a_2+b_2 )+i\,\omega\,(a_2-b_2)-2a_2b_2}{i\,\kappa\,(a_2+b_2 )+i\,\omega\,(a_2-b_2)+2a_2b_2}.$$
On the other hand, setting $a_1=a_2=b_1=b_2=0$ the transmission matrix becomes
$$T=\frac{1}{f}\,\left(\begin{array}{cc}
a &i\,c\\
i\,c & a
\end{array}\right),$$
with
\begin{eqnarray*}
a&=&(\alpha\,\kappa+\alpha\,\omega+\beta\,\kappa-\beta\,\omega-2\alpha\,\beta)\,(\alpha\,\kappa+\alpha\,\omega+\beta\,\kappa-\beta\,\omega+2\alpha\,\beta),\\
c&=&4\alpha\,\beta\,(\alpha\,\kappa+\alpha\,\omega+\beta\,\kappa-\beta\,\omega),\\
f&=&(i\alpha\,\kappa+i\alpha\,\omega+i\beta\,\kappa-i\beta\,\omega-2\alpha\,\beta)\,(i\alpha\,\kappa+i\alpha\,\omega+i\beta\,\kappa-i\beta\,\omega+2\alpha\,\beta),
\end{eqnarray*}
which simplifies further to
\begin{equation}
T=\frac{1}{(i\kappa-\alpha)\,(i\kappa+\alpha)}\,\left(\begin{array}{cc}
(\kappa-\alpha)\,(\kappa+\alpha) &i\,2\kappa\,\alpha\\
i\,2\kappa\,\alpha & (\kappa-\alpha)\,(\kappa+\alpha)
\end{array}\right),
\end{equation}
by setting $\alpha=\beta.$

\end{document}